\documentclass{jpp}

\usepackage[utf8]{inputenc}
\usepackage[T1]{fontenc}
\usepackage{newtxtext,newtxmath}

\usepackage{graphics}      
\usepackage{graphicx}      
\usepackage{longtable}     
\usepackage{url}           
\usepackage{amsmath, amssymb, amsbsy}
\usepackage{natbib}
\usepackage{bm}            
\usepackage{dcolumn}       
\usepackage{xcolor}		   
\usepackage[colorlinks=true,allcolors=blue,breaklinks=true]{hyperref} 
\usepackage[normalem]{ulem} 
\usepackage{appendix}
\usepackage{lipsum}
\usepackage{apptools}
\usepackage[linesnumbered,ruled,vlined]{algorithm2e}
\graphicspath{{./}{./figs/}}

\renewcommand{\vec}[1]{\boldsymbol{#1}}

\newcommand{\oldhat}[1]{\hat{#1}}
\let \oldhat \hat
\renewcommand{\hat}[1]{\oldhat{\boldsymbol{#1}}}

\newcommand{\overbar}[1]{\mkern 1.5mu\overline{\mkern-1.5mu#1\mkern-1.5mu}\mkern 1.5mu}

\shorttitle{Quasilinear gyrokinetic theory: A derivation of QuaLiKiz}
\shortauthor{C. D. Stephens, X. Garbet, J. Citrin, C. Bourdelle, K. L. van de Plassche, and F. Jenko}

\title{Quasilinear gyrokinetic theory: \\A derivation of QuaLiKiz}
\author{C. D. Stephens\aff{1,2,} \corresp{\email{cdstephens@.ucla.edu}}, X. Garbet\aff{3}, J. Citrin\aff{4}, C. Bourdelle\aff{3}, K. L. van de Plassche\aff{4}, and F. Jenko\aff{2}}
\affiliation{
	\aff{1} University of California, Los Angeles, 475 Portola Plaza, Los Angles, California 90095, USA
	\aff{2} Max Planck Institute for Plasma Physics, Boltzmannstr. 2, 85748 Garching, Germany 
	\aff{3} CEA, IRFM, F-13108 Saint Paul-lez-Durance, France
	\aff{4} DIFFER—Dutch Institute for Fundamental Energy Research, De Zaale 20, 5612 AJ Eindhoven, \\The Netherlands
}

\begin{document}

	\maketitle
\begin{abstract}
	In order to predict and analyze turbulent transport in tokamaks, it is important to model transport that arises from microinstabilities. For this task, quasilinear codes have been developed that seek to calculate particle, angular momentum, and heat fluxes both quickly and accurately. In this tutorial, we present a derivation of one such code known as QuaLiKiz, a quasilinear gyrokinetic transport code. The goal of this derivation is to provide a self-contained and complete description of the underlying physics and mathematics of QuaLiKiz from first principles. This work serves both as a comprehensive overview of QuaLiKiz specifically as well as an illustration for deriving quasilinear models in general. 
\end{abstract}

\section{Introduction}

The development of tractable transport models is crucial to further the study and operation of tokamaks. Accurately characterizing the particle, angular momentum, and heat transport in the tokamak core requires the understanding of turbulence driven by microinstabilities, as these instabilities drive much of the particle, momentum, and heat transport in the core. Integrated modeling codes seek to predict and simulate tokamak discharges via the inclusion of various different physics and sources, including from microinstabilities. Nonlinear simulations of the kinetic equations are the most accurate way to compute the transport from microinstabilities. For reference, the cost of such a nonlinear simulation is on the order of $10^4$ CPUh to $10^5$ CPUh at a single radial point, while integrated modeling frameworks require thousands flux calculations for every second of a plasma discharge in a large tokamak device \citep{citrin2017}. Multi-scale simulations that take into account the interplay of instabilities across wide ranges of time scales are even more expensive \citep{waltz2007, goerler2008, howard2016}. Even linear kinetic simulations can prove to be intractable for integrated modeling if not reduced enough. Thus, it is imperative to develop and refine kinetic models that are both accurate enough to account for transport from microinstabilities and fast enough to be coupled to an integrated modeling framework. 

QuaLiKiz is a quasilinear gyrokinetic transport model originally based on the linear eigenvalue code Kinezero. Pieces of the derivation have been published throughout the years including in \citet{bourdelle2000}, \citet{bourdelle2002}, and \citet{bourdelle2007}. Th underlying principles of the code regarding the variational and action-angle approaches can be traced to \citet{garbet1990}, and upgrades to the physics including angular momentum transport \citep{cottier2014} and numerical improvements \citep{citrin2017} have been made since its original development. The goal of QuaLiKiz is to calculate the quasilinear transport that arise from microinstabilities. The core principle is to develop linearize the kinetic equations and solve the dispersion relation to find the complex frequencies for microinstabilities, namely the ion temperature gradient (ITG), electron temperature gradient (ETG), and trapped electron mode (TEM) instabilities. Upon solving the linear problem, we then incorporate nonlinear physics to compute particle, angular momentum, and heat fluxes. We do so via a quasilinear approach by coupling the linear characteristics of the problem together and using previously performed nonlinear kinetic simulations to saturate the perturbed state. Thus, while the amplitudes of the modes are set by nonlinear physics, the key transport features can be constructed from the linear regime. Quasilinear methods have been shown to be valid in the tokamak core. Moreover, the quasilinear codes are much faster than fully nonlinear kinetic codes. QuaLiKiz in particular can perform a full computation in $\sim 1$ CPUs per wavenumber \citep{citrin2017}.

As a gyrokinetic code, QuaLiKiz is well suited to model the core of tokamak devices which are strongly magnetized. Gyrokinetics is a popular approach to investigate turbulent phenomena in magnetized plasmas such as those of fusion devices \citep{brizard2007gyrokinetic, cary2009guiding}. Gyrokinetics is well suited in scenarios where the microscopic dynamics are subject to the gyrokinetic ordering. Essentially, we apply gyrokinetics to situations where we can decouple the fast gyromotion of the charged particle from the slow drift motion; this can be done when the time scale of the gyromotion is significantly faster than all other time scales in the system and when the gyroradius is smaller than almost all other length scales in the system. In such a scenario, the magnetic moment is conserved, leading to a significant reduction in the complexity of the dynamics \citep{stephens2017moment}. Moreover, gyrokinetics incorporates an ordering where the modes are anisotropic and flute-like, meaning that the characteristic parallel wavelength of the mode is large but perpendicular wavelengths can be comparable to the gyroradius. Thus, gyrokinetics is well suited for theoretical and quantitative investigations of magnetized plasma microturbulence. As a result, gyrokinetics has been used and applied in a wide variety of systems \citep{Wan, Rogers, Wang, Pueschel1, Pueschel2, Howes1, Howes2, Told1, Navarro, Told2}. Even beyond tokamaks, progress is being made in simulating stellarator plasmas in gyrokinetic codes \citep{jenko2002, xanthopoulos2007, mynick2010, nunami2010, baumgaertel2011}. QuaLiKiz in particular, however, assumes an axisymmetric geometry to simplify the dynamics, meaning QuaLiKiz is only suitable for tokamaks and not stellarators.

Aside from the well established gyrokinetic approach, the key assumption behind QuaLiKiz is the quasilinear approximation. In nonlinear simulations, turbulent fluctuations eventually saturate due to coupling mechanisms between different modes. However, it has been found that the nonlinear mode structure can resemble the underlying linear mode structure; in particular, the cross phases between fluctuating quantities in nonlinear simulations are identical to that of linear simulations \citep{dannert2005, jenko2005}. In such situations, one also finds that ratios of the particle and heat fluxes calculated in the linear regime match those calculated in the nonlinear regime and that the real part of the nonlinear mode frequency resembles that of the linear mode \citep{merz2008, goerler2008}. Moreover, it has been found that when different instabilities are found in the linear regime, their interplay can manifest in the nonlinear regime \citep{merz2010}. This motivates a quasilinear approach where the equilibrium distribution function slowly evolves in comparison to the time scale of the instability, essentially taking a mean field theory approach. Then, the linear response is acquired and used to inform the the first order nonlinear behavior of the system. Quasilinear flux ratios are then calculated and each flux is appropriately saturated to the correct magnitude using a nonlinear saturation rule informed by nonlinear physics. The approach allows us to exploit the fact that the nonlinear state resembles the linear state to perform flux calculations without needing to carry out a full nonlinear simulation \citep{citrin2012}.

However, constructing a quasilinear code instead of a nonlinear code is alone not enough to increase the speed of calculations. Rather, a litany of approximations and reductions are necessary. Aside from other typical approximations for gyrokinetic tokamak codes (e.g. nonrelativistic particles, quasineutrality), QuaLiKiz makes use of the following assumptions:
\begin{itemize}
	
	\item Adiabatic invariance. By exploiting the adiabatic invariants of the system, we can formulate the Vlasov equation with action-angle variables. This requires that the single-particle Hamiltonian be slowly varying in time in comparison to the characteristic frequencies of motion. These frequencies correspond to the cyclotron motion, the bounce-transit motion, and the toroidal drift and precession.
	
	\item Shifted Maxwellian with low Mach number and the $\delta f$ approximation. QuaLiKiz linearizes the Vlasov equation by assuming a small perturbation from the shifted Maxwellian. Although we include the effect of bulk plasma rotation, we operate in the limit that the Mach number associated with the rotation is small. 
	
	\item Electrostatic fluctuations. The code allows for electrostatic perturbations and an equilibrium electric field. The absence of magnetic perturbations allows for the exclusive use of Poisson's equation while neglecting Ampere's law, thus simplifying the linear problem. To simplify the guiding center motion, we require that the equilibrium electrostatic potential is small compared to the characteristic thermal energy.
	
	\item Trapped electron collisions. As an approximation, we utilize a Krook collision operator for trapped electrons and neglect collisions entirely for passing electrons and all ions.	
	
	\item Shifted circle geometry with small inverse aspect ratio. This simplified geometry is used to calculate the magnetic drifts and perform integrals over the pitch angle with ease. The $s-\alpha$ model gives rise to a radial shift in the concentric flux surfaces called the Shafranov shift. The effect of this shift is included when calculating the magnetic drifts, but ignored when considering the bounce-transit motion. Thus, the treatment of guiding center motion with respect to the geometry is inherently inconsistent. Moreover, the $s-\alpha$ model is ad-hoc and does not solve the Grad-Shafranov equation.
	
	\item Gaussian eigenfunctions. Instead of using a self-consistent eigenfunction for the electrostatic modes, QuaLiKiz assumes the modes take the form of a Gaussian. The shift and width of the Gaussian are calculated in the high mode frequency limit as functions of the mode frequency, and substituted back into the dispersion relation. 
	
	\item Strong ballooning. The electrostatic modes are assumed to be heavily localized around their rational flux surface. This allows for a Fourier link between the minor radius $r$ and the poloidal angle $\theta$, thus simplifying the calculation. The localization also creates a separation of scales, thus allowing the integrals to be more easily approximated. 
	
	\item Strongly passing and strongly trapped particles. Trapped and passing particles are considered to be respectively strongly trapped and strongly passing. For trapped particles, this greatly simplifies the relation between the physical toroidal and poloidal angles and the action angles and leads to a kinetic bounce average that is similar to the gyro-average. For passing particles, the strongly passing assumption simplifies the integrals over the pitch angle due to the dominating parallel velocity.

\end{itemize}

The goal of this work is to derive the analytic equations for QuaLiKiz step by step. Although various overviews of the QuaLiKiz and Kinezero framework have already been published \citep{bourdelle2015, bourdelle2016, citrin2017}, no combination of currently published works derive the entirety of the model from first principles. We seek to fill this gap by offering a comprehensive and complete formulation of QuaLiKiz. This work will then as a result serve as a guide for improving upon QuaLiKiz and attaining physical and mathematical intuition as to its key principles, approximations, and computational methods. In addition, we also outline the new computational method used to numerically calculate 1-dimensional and 2-dimensional integrals. Moreover, this sort of work serves as a tutorial for those seeking to understand the fundamental considerations in the formulation of any quasilinear tokamak code. While many codes offer comprehensive manuals and describe the key principles at play, the process of creating such a code from scratch can often appear opaque and unintuitive. Thus, this derivation also serves as a tutorial for those who seek to understand the physical, mathematical, and computational aspects of quasilinear modeling in all their gory details.

The paper is organized as follows: Section~\ref{Action Angle Variables} reviews the action-angle formalism and derives explicit expressions for the action-angle variables from physical variables. In Section~\ref{The Vlasov Equation}, we linearize the Vlasov equation and expand the perturbed distribution function and electrostatic potential using a Fourier series to derive the dispersion relation. To solve the dispersion relation, we must integrate over all of phase space, resulting in a functional that depends on the complex frequency of the mode. Section~\ref{Ballooning} examines the ballooning transform and its role in simplifying the dispersion relation as well as the characteristics of the electrostatic perturbation. Sections~\ref{Adiabatic}~-~\ref{Passing} calculate the adiabatic, trapped, and passing parts of the functional, respectively, resulting in a reduced expression for the dispersion relation. Section~\ref{Quasilinear} applies these results to the quasilinear problem to derive expressions for the particle, toroidal angular momentum, and heat fluxes. Section~\ref{Saturation} connects the quasilinear results with nonlinear physics with the use of a saturation rule. Section~\ref{Numerical} explains the method of contour integration used in QuaLiKiz to find the eigenmodes and the newly implemented numerical integration method based on the Genz and Malik algorithm \citep{genz1980}. Finally, we summarize our work Section~\ref{Conclusions}. We include Appendix~\ref{Fried and Conte} to serve as a brief explanation of Fried and Conte integrals. In addition, we derive the magnetic drift velocity in an $s-\alpha$ equilibrium in Appendix~\ref{Magnetic} and briefly discuss the inclusion trapped electron collisions in Appendix~\ref{Collisions}. The derivation is performed in SI units, and we set the Boltzmann constant $k_B = 1$ such that our temperatures are in units of energy. 

\section{Action Angle Variables}\label{Action Angle Variables}
We first restrict ourselves to the collisionless Vlasov equation. Since the inclusion of collisions do not affect the fundamental approach, we examine them later in Appendix~\ref{Collisions}. The Vlasov equation is
\begin{equation}
\frac{\partial f}{\partial t} + \left\{f, H\right\} = 0,
\end{equation}
where $f$ is the distribution function, $H$ is the single particle Hamiltonian, and $\left\{\cdot, \cdot\right\}$ denotes the Poisson bracket. Using phase space coordinates, this can be written as
\begin{equation}
\frac{\partial f}{\partial t} + \dot{\vec{q}} \cdot \frac{\partial f}{\partial \vec{q}} + \dot{\vec{p}} \cdot \frac{\partial f}{\partial \vec{p}} = 0,
\end{equation}
where, for a single particle $\vec{q}$ is the position, $\vec{p}$ is the canonical momentum, and the time derivatives are given by Hamilton's equations of motion. For electromagnetic fields relevant to a tokamak, the Hamiltonian of a single charged particle is non-trivial. Although this form of the Vlasov equation and others like it offer an intuitive physical picture, these coordinates can make solving the system quite cumbersome. QuaLiKiz instead employs an action-angle formalism to simplify the perturbative analysis. Such a formalism in the context of tokamak physics was first elaborated by \citet{kaufman1972} and expanded upon by \citet{mahajan1985}. The core principle is to define a canonical transformation,
\begin{equation}
(\vec{q}, \vec{p}) \to (\bm{\alpha}, \vec{J}),	
\end{equation}
for which Hamilton's equations of motion simplify in the new phase space ($\bm{\alpha}, \vec{J}$). By restricting ourselves to a canonical transformation, we preserve the form of Vlasov's equation. The coordinates $\bm{\alpha}$ and $\vec{J}$ respectively correspond to the action angles and adiabatic invariants of our system. It is well known \citep{goldstein2001classical} that Hamilton's equations of motion then reduce to
\begin{align}
\frac{\partial H}{\partial \bm{\alpha}} &= - \dot{\vec{J}} = \vec{0},\\
\frac{\partial H}{\partial \vec{J}} &= \dot{\bm{\alpha}} = \vec{\Omega},
\end{align}
where $\vec{\Omega}$ are the constant angular frequencies associated with each adiabatic invariant. At first glance, it may seem that we have simply shifted the difficulty of the problem to calculating this new canonical transformation itself. The power of this method comes from analyzing the unperturbed system and then including electromagnetic fluctuations in the Hamiltonian. 

We define the Hamiltonian to be
\begin{equation}
H = H_0 + \delta h,
\end{equation}
where the unperturbed Hamiltonian is simply
\begin{equation}
H_0 = \frac{1}{2m} \left(\vec{p}^2 - e \vec{A}_0^2\right) + e \Phi.
\end{equation}
Here, $m$ and $e$ are respectively the mass and charge of the particle, $\vec{A}_0$ is the equilibrium vector potential, and $\Phi$ is the equilibrium electrostatic potential. Since QuaLiKiz operates in the electrostatic limit, we therefore apply a perturbation $\delta h$ such that
\begin{equation}
\delta h = e \phi,
\end{equation}
where $\phi$ is the electrostatic perturbation. We then define the action-angle coordinates in reference to the unperturbed Hamiltonian,
\begin{align}
\frac{\partial H_0}{\partial \bm{\alpha}} &= \vec{0},\\
\frac{\partial H_0}{\partial \vec{J}} &= \vec{\Omega}.
\end{align}
Hamilton's equations of motion then become
\begin{align}
\frac{\partial H}{\partial \bm{\alpha}} & = - \dot{\vec{J}} = e \frac{\partial \phi}{\partial \bm{\alpha}},\\
\frac{\partial H}{\partial \vec{J}} & = \dot{\bm{\alpha}} = \vec{\Omega} + e \frac{\partial \phi}{\partial \vec{J}}.
\end{align}
We note that because the unperturbed Hamiltonian is a function of $\vec{J}$ and not $\bm{\alpha}$, all equilibrium quantities are also only functions of $\vec{J}$. Furthermore, any function of $\bm{\alpha}$ is periodic with respect to $\bm{\alpha}$; thus, the perturbed quantities in our system admit a Fourier series expansion. Moreover, it can be shown that $\bm{\alpha}, \vec{J}$ are canonical coordinates even after introducing a perturbation \citep{mahajan1985}. These features will simplify the derivation greatly. 

The next task is to define the canonical transformation by specifying the action-angle variables in terms of the position $\vec{r}$ and the velocity $\vec{v}$ of the particle. The three adiabatic invariants in a tokamak correspond to the magnetic moment, the longitudinal invariant (also known as the bounce-transit action), and the poloidal flux. They are defined as follows:
\begin{align}
&J_1 = \frac{m}{e} \mu,\\
&J_2 = \frac{1}{2 \pi} \oint m p_\parallel dl = \frac{1}{2 \pi} \oint \left(m v_\parallel + e A_\parallel\right) dl,\\
&J_3 = P_{\varphi} = \frac{m v_\parallel R_0 B_\varphi^0}{B} - e \psi.
\end{align}
Here, $\mu = W_\perp/B$ is the magnetic moment, where $W_\perp = \frac{1}{2} m v_\perp^2$ is the kinetic energy associated with the velocity perpendicular to the magnetic field $\vec{B}$. Meanwhile, $v_\parallel$ and $A_\parallel$ are the components of the velocity and vector potential parallel to the magnetic field, respectively, with $dl$ being the signed differential length along the particle orbit. We also define $\psi$ to be minus the poloidal magnetic flux normalized to $2 \pi$, which is calculated by integrating the flux of the magnetic field through a disk tangent to the flux surface everywhere:
\begin{equation}
	\psi = - \frac{1}{2\pi} \int_S \vec{B} \cdot d\vec{S}. 
\end{equation}
 The following subsections discuss each of the three adiabatic invariants and define their associated action angles and angular frequencies. For the remainder of the derivation, we also use the spatial coordinates $\vec{r} = (r, \theta, \varphi)$, where $r$ is the minor radial position, $\theta$ is the geometric poloidal angle, and $\varphi$ is the geometric toroidal angle. We use a right-handed coordinate system such that $\hat{r} \times \hat{\theta} = \hat{\varphi}$. For further references characterizing the action angles $\bm{\alpha}$, we refer the reader to Garbet's work in \citet{garbet2001} and \citet{garbet1990}. 

\subsection{Magnetic Moment}
In the presence of a magnetic field, charged particles gyrate about the field line at the cyclotron frequency $\Omega_1 = eB/m$. With a strong enough magnetic field, the cylcotron frequency is much larger than any other characteristic frequency in the plasma. Under such conditions, the magnetic moment $\mu$ is adiabatically conserved \citep{kruskal1962asymptotic, banos1967guiding, brizard2007gyrokinetic, cary2009guiding, stephens2017moment}, and the gyromotion can be decoupled from the guiding center motion: 
\begin{align}
&r = r_G + \rho \cos(\alpha_1),\\
&\theta = \theta_G + \frac{\rho}{r} \sin(\alpha_1),\\
&\varphi = \varphi_G,
\end{align}
where $\alpha_1$ is equivalent to the gyrophase, $\rho$ is the gyroradius, and the subscript ``G'' refers to the location of the particle's guiding center. These guiding center variables obey the guiding center equations of motion. Ordinarily, the exact invariant associated with the gyromotion depends on the electrostatic potential. For QuaLiKiz, we assume that the electrostatic field is small compared to the kinetic energy. Thus, we simply take $J_1$ to be the ordinary magnetic moment $\mu = W_\perp/B$. 

Later in the derivation, we will need to take average various functions over the gyrophase $\alpha_1$ by integrating over $\alpha_1$. We therefore consider the general integral
\begin{equation}
g_{n_1} = \int_{-\pi}^{\pi} \frac{d \alpha_1}{2 \pi} g(\vec{r}) e^{- i n_1 \alpha_1},
\end{equation}
where $n_1$ is an integer. It will be shown that later that factors of $e^{- i n_1 \alpha_1}$ arise from taking Fourier expansions in terms of $\alpha_1$. We define the Fourier transform of $g$ to be
\begin{equation}
\tilde{g}(\vec{k}) = \int d^3r g(\vec{r}) e^{i \vec{k}\cdot \vec{r}},
\end{equation}
with the corresponding inverse Fourier transform
\begin{equation}
g(\vec{r}) = \int \frac{d^3k}{(2\pi)^3} \tilde{g}(\vec{k}) e^{-i \vec{k} \cdot \vec{r}}. 
\end{equation}
We use the Fourier transform to obtain
\begin{equation}
g_{n_1} = \int_{-\pi}^{\pi} \frac{d \alpha_1}{2 \pi} \int \frac{d^3k}{(2\pi)^3}\tilde{g}(\vec{k})e^{-i \vec{k} \cdot \vec{r}- i n_1 \alpha_1} = \int_{-\pi}^{\pi} \frac{d \alpha_1}{2 \pi} \int \frac{d^3k}{(2\pi)^3}\tilde{g}(\vec{k})e^{-i \vec{k} \cdot \bm{\rho} - i n_1 \alpha_1} e^{-i \vec{k} \cdot \vec{R}_G}. 
\end{equation}
Here, we have decoupled the gyromotion from the guiding center motion via $\vec{r} =\vec{R}_G + \bm{\rho}$. We then write
\begin{equation}
\vec{k} \cdot \bm{\rho} = k_\perp \rho \cos(\alpha_1),
\end{equation}
where 
\begin{equation}
k_\perp = \left|\vec{k} - \vec{k} \cdot \hat{b}\right| \approx \sqrt{k_r^2 + k_\theta^2}. 
\end{equation}
Note that according to our definition of the Fourier transform, $k_r$ and $k_\theta$ are operators in real space such that
\begin{align}
k_r &\to i \frac{\partial}{\partial r},\\
k_\theta & \to \frac{i}{r} \frac{\partial}{\partial \theta}.
\end{align}
We may then integrate over $\alpha_1$ independently, leading to
\begin{equation}
\int^\pi _{-\pi}  \frac{d \alpha_1}{2 \pi} e^{- i k_\perp \rho \cos(\alpha_1) - i n_1 \alpha_1} = (-i)^{n_1} J_{n_1}(k_\perp \rho),
\end{equation}
where $J_n$ is the $n$th Bessel function of the first kind. Therefore, we finally have that
\begin{equation}
g_{n_1} = \int \frac{d^3 k}{(2 \pi)^3} (-i)^{n_1} J_{n_1}(k_\perp \rho) \tilde{g}(\vec{k}) e^{- i \vec{k} \cdot \vec{R}_G} = (-i)^{n_1} \left(J_{n_1}(k_\perp \rho) \cdot g\right)(\vec{R}_G). 
\end{equation}
As a shorthand, we treat the Bessel function in real space as a differential operator that acts on $g$, after which we evaluate the resulting function at the guiding center. The Bessel function is simply a scalar function in Fourier space instead of a differential operator. The case of $n_1 = 0$ corresponds to the well known gyro-average. After completing the gyro-average, all functions are evaluated at the guiding center. Thus, we drop the subscript ``G" for convenience and treat all spatial variables as those corresponding to the guiding center. The adiabatic invariants $J_2$ and $J_3$ are explicitly calculated within the guiding center framework where we hold $\mu$ constant and ignore the cyclotron motion. 

\subsection{Longitudinal Invariant}
To calculate $J_2$, we consider the guiding center particle motion along a magnetic field line; such a particle completes bounce-transit orbits with frequency $\Omega_2$. This is the bounce frequency for trapped particles and the transit frequency for passing particles. Here, we neglect excursions from the field line due to various guiding center drifts by holding $r$ constant. For an extended treatment of bounce-transit motion, see \citet{brizard2011compact} and \citet{stephens2020}. 

Assuming that the equilibrium electrostatic potential is small, the guiding center velocity parallel to the magnetic field is
\begin{equation}
\left|v_\parallel\right| = \sqrt{\frac{2}{m}(E - \mu B)},
\end{equation}
where $E$ is the total kinetic energy of the particle. As an approximation, we take the typical equilibrium magnetic field to be of the form 
\begin{equation}
\vec{B} = B_\varphi(r, \theta) \hat{\bm{\varphi}} + B_\theta(r, \theta) \hat{\bm{\theta}} = \frac{1}{1 + r/R_0 \cos(\theta)}\left( B^0_\varphi(r) \hat{\bm{\varphi}} + B^0_\theta(r) \hat{\bm{\theta}}\right),
\end{equation}
where $R_0$ is the major radius. This corresponds to the magnetic field in a circular-cross section tokamak without any Shafranov shift. Defining the inverse aspect ratio $\epsilon = r/R_0$, we recognize that this circular equilibrium is the small $\epsilon$ limit of a more general axisymmetric equilibrium. QuaLiKiz is thus well suited to machines where the aspect ratio of the device is $\sim 3$ or larger. Devices which smaller aspect ratios such as spherical tokamaks, however, cannot be reliably simulated in QuaLiKiz.

 A particle is considered trapped if it reflects at a bounce angle $\theta_b$, which requires
\begin{equation}
\frac{\mu B^0 (r)}{E} \ge 1 - \epsilon.
\end{equation}
 Otherwise, the particle is considered passing since it will simply continue traveling along the magnetic field line without reflecting. We rewrite $v_\parallel$ to be
\begin{equation}
v_\parallel = \sqrt{\frac{2 T}{m}} \epsilon_\parallel \sqrt{\xi} \sqrt{1 - \lambda b(r, \theta)}.
\end{equation}
Here, $\xi = E/T$ where $T$ is the temperature and $\epsilon_\parallel = \pm 1$ determines the sign of the parallel velocity. We also define
\begin{align}
\lambda &= \frac{\mu B\left(r, \theta = 0\right)}{E} = \frac{v_\perp^2}{v^2 b(r, \theta)},\\
b(r, \theta) &= \frac{B(r, \theta)}{B(r, \theta = 0)}. 
\end{align}
It is clear then that $\lambda$ is a pitch angle parameter and determines whether the particle is trapped or passing.

The bounce-transit frequency is defined as
\begin{equation}
\left|\Omega_2\right| = \frac{2 \pi}{T_2},
\end{equation}
where
\begin{equation}
T_2 = \oint \frac{d\theta}{\left|\frac{d\theta}{dt}\right|}. 
\end{equation}
We note that one full poloidal orbit for trapped particles includes both the forward motion, where $\theta$ goes from $-\theta_b$ to $\theta_b$, and the backward motion, where $\theta$ goes from $\theta_b$ to $-\theta_b$. For passing particles, the poloidal orbit only includes one full pass where $\theta$ goes from $-\pi$ to $\pi$. The sign of the transit frequency for passing particles is aligned with that of the parallel velocity and is thus determined by $\epsilon_\parallel$, while the bounce frequency is always positive for trapped particles. Assuming that $B_\varphi \gg B_\theta$, then $\hat{b}$, the direction of the magnetic field, is approximately $\hat{\bm{\varphi}}$. We again emphasize that this approximation breaks down for devices such as spherical tokamaks. Therefore, we write 
\begin{equation}
v_\parallel = \vec{v}\cdot\hat{b} \approx \dot{\varphi} \left(R_0 + r \cos(\theta)\right) \approx q R_0 \dot{\theta}.
\end{equation}
Here, we have defined the safety factor
\begin{equation}
q(r) = \frac{d \varphi}{d \theta} \approx \frac{r B_\varphi}{R_0 B_\theta}.
\end{equation}
The safety factor describes how many times a magnetic field line wraps around toroidally per poloidal turn. The magnitude of the bounce-transit frequency is then
\begin{equation}
\left|\Omega_2\right| = \sqrt{\frac{2T}{m}} \frac{\sqrt{\xi}}{qR_0} \overbar{\Omega}_2 (r, \lambda),	
\end{equation}
where we define
\begin{equation}
\overbar{\Omega}_2 (r, \lambda) = \frac{2 \pi}{\oint d\theta \frac{1}{\sqrt{1 - \lambda b(r, \theta)}}}.
\end{equation}
Note that for passing particles, we take the sign of the transit frequency to be the sign of the parallel velocity and multiply by $\epsilon_\parallel$ accordingly. We then calculate $\overbar{\Omega}_2$ in the small $\epsilon$ limit to
\begin{equation}
\overbar{\Omega}_2 \approx
	\left\{\begin{aligned}
	& \frac{\pi \sqrt{\epsilon}}{2 \sqrt{2}  K(\kappa)} &&\text{ if } 0\le \kappa < 1 \text{ (trapped)},\\
	& \frac{\pi \kappa \sqrt{\epsilon}}{\sqrt{2} K\left(\kappa^{-1}\right)} &&\text{ if } 1 < \kappa < \infty \text{ (passing)}.
	\end{aligned} \right. 
\end{equation}
Here, $K$ is the complete elliptic integral of the first kind and $\kappa$ is a trapped parameter defined such that
\begin{equation}
\lambda = 1 - 2 \epsilon \kappa^2.
\end{equation}
In the small $\epsilon$ limit, $0 \leq \kappa < 1$ for trapped particles and $1 < \kappa < \infty$ for passing particles. We also calculate the bounce-transit action to be
\begin{equation}
J_2 \approx \left\{\begin{aligned} 
& \frac{8 m q R_0 \sqrt{E/m} \sqrt{\epsilon}}{\pi} \left(E\left(\kappa\right) - \left(1-\kappa^2\right) K\left(\kappa\right)\right) &&\text{ if } 0\le \kappa < 1 \text{ (trapped)},\\
& \frac{4 m q R_0 \sqrt{E/m} \sqrt{\epsilon}}{\pi} \kappa E\left(\kappa^{-1}\right) + e \Phi_t &&\text{ if }  1 < \kappa < \infty \text{ (passing)},
	\end{aligned}\right. 
\end{equation}
where $E$ is the complete elliptic integral of the second kind and $\Phi_t$ is the toroidal flux normalized by $2 \pi$. The flux term is absent for trapped particles since the closed line integral of $A_\parallel$ is zero for trapped orbits. 

Calculating the angular variable $\alpha_2$ requires the explicit equation of motion
\begin{equation}
	\frac{d \alpha_2}{dt} = \Omega_2.
\end{equation}
This is of course the definition of $\alpha_2$ such that it is conjugate to the action variable $J_2$. To find an explicit expression for $\alpha_2$ in terms of the poloidal angle $\theta$, we make use of the chain rule,
\begin{equation}
\frac{d\alpha_2}{dt} = \frac{d\alpha_2}{d\theta} \dot{\theta} = \frac{d\alpha_2}{d\theta} \frac{v_\parallel}{q R_0}  = \Omega_2.
\end{equation}
We emphasize that $\Omega_2$ is not dependent on $\alpha_2$ or $\theta$. Thus, this differential equation can be integrated using elliptic functions, leading to an expression of $\alpha_2$ in terms of $\theta$. We use the convention that $\alpha_2(\theta = 0) = 0$, leading to
\begin{equation}
	\alpha_2 = \int^{\theta}_{0} d \theta' \frac{q R_0 \Omega_2}{v_{\parallel}} = \int^{\theta} d \theta' \frac{q R_0 \Omega_2}{\sqrt{2T / m} \epsilon_{\parallel} \sqrt{\xi} \sqrt{1 - \lambda b}}. 
\end{equation}
For trapped particles, we must keep in mind that $\epsilon_{\parallel}$ switches sign after the particle bounces. The integral can then be simplified in the small $\epsilon$ limit, leading to 
\begin{equation}
\alpha_2 \approx \left\{ \begin{aligned} & \frac{\pi}{2} \frac{F\left(\sin^{-1}\left[\kappa \sin\left(\frac{\theta}{2}\right)\right], \kappa\right)}{K\left(\kappa\right)} && \text{ if } 0\le \kappa < 1 \text{ (trapped)},\\
& \pi \frac{F\left(\frac{\theta}{2}, \kappa^{-1}\right)}{K\left(\kappa^{-1}\right)} &&\text{ if } 1 < \kappa < \infty \text{ (passing)},
\end{aligned} \right. 
\end{equation}
where $F$ is the incomplete integral of the first kind. Essentially, the integral takes the same form as when calculating $\Omega_2$, the primary difference being that we integrate up to arbitrary $\theta$ rather than up to the bounce angle $\theta_b$ for trapped particles or up to $\pi$ for passing particles.

Finally, let $G \left(\epsilon_\parallel, \theta \right)$ be a quantity that varies over the bounce-transit orbit along the field line. It is often of interest to time average $G$ over the orbit; we define the bounce-transit average $\left<G \left(\epsilon_\parallel, \theta \right)\right>$ to be
\begin{equation}
\left<G \left(\epsilon_\parallel, \theta \right)\right> = \frac{1}{T_2} \oint \frac{d\theta}{\left|\frac{d\theta}{dt}\right|} G \left(\epsilon_\parallel, \theta \right) = \frac{\oint d\theta \frac{G \left(\epsilon_\parallel, \theta \right)}{\sqrt{1 - \lambda b}}}{\oint d\theta \frac{1}{\sqrt{1-\lambda b}}}. 
\end{equation}
For passing particles, the average is explicitly
\begin{equation}
\oint d\theta \frac{G \left(\epsilon_\parallel, \theta \right)}{\sqrt{1-\lambda b}} = \int_{-\pi} ^\pi d\theta \frac{G \left(\epsilon_\parallel, \theta \right)}{\sqrt{1-\lambda b}},
\end{equation}
while for trapped particles the average is instead
\begin{equation}
\oint d\theta \frac{G \left(\epsilon_\parallel, \theta \right)}{\sqrt{1-\lambda b}} = \int_{-\theta_b} ^{\theta_b} d\theta \frac{G\left(\epsilon_\parallel, \theta\right) + G\left(-\epsilon_\parallel, \theta\right)}{\sqrt{1-\lambda b}}.
\end{equation}
Note that because the line integral must be closed, a sum over $\epsilon_\parallel$ must be performed for trapped particles so that quantities such as $v_\parallel$ average to 0. 

In this discussion so far, we have neglected any magnetic drifts and excursions from the field line. We include such effects in the next section, as they characterize the third adiabatic invariant --- the poloidal flux.

\subsection{Poloidal Flux}
In an axisymmetric equilibrium, the canonical toroidal momentum, $P_\varphi$, is conserved since no external quantities depend explicitly on the toroidal angle $\varphi$. From guiding center theory, we can write the canonical toroidal momentum as 
\begin{equation}
P_\varphi = \frac{m v_\parallel R_0 B_\varphi^0}{B} - e \psi. 
\end{equation}
This is an exact invariant of the system. We construct $J_3$ such that it approximates $P_\varphi$ provided that the poloidal flux term dominates. For typical parameters in a tokamak plasma this is indeed the case, since $\left(P_\varphi + e \psi \right) / \left(e \psi \right) \sim \sqrt{m T}/(eB R_0)$. Inputting JET-like parameters, $T = 5$ keV, $m = m_D, B = 3$ T, $R_0 = 3$ m, then $\sqrt{m T}/(eB R_0) \sim 10^{-3}$, making this a very reasonable approximation. We therefore write
\begin{equation}
J_3 = - e \psi.
\end{equation}
To calculate the poloidal flux, we utilize Stoke's theorem; the surface integral of $\vec{B$} simply becomes a closed line integral of $\vec{A}$ to find
\begin{equation}
A_\varphi(r, \theta) = \frac{1}{1 + \epsilon \cos(\theta)} \int^r B_\theta(r^\prime, \theta) (1 + r^\prime/R_0 \cos(\theta)) dr^\prime = \frac{\int^r B^0_\theta(r^\prime) dr^\prime}{1 + \epsilon \cos\left(\theta\right)}.  
\end{equation}
Thus, we obtain
\begin{equation}
J_3 = - e \psi = - \frac{1}{2\pi} \int_0 ^{2\pi}  e(1 + \epsilon \cos(\theta)) A_\varphi d\varphi = e \int^r R_0 B_\theta^0(r^\prime) dr^\prime.
\end{equation}
We see then that $J_3$ is purely a function of $r$ such that
\begin{equation}
	\frac{d J_3}{dr} \approx -\frac{e r B}{q}.
\end{equation}

We next calculate $\Omega_3$, which is the toroidal precession frequency for trapped particles and the toroidal rotation frequency for passing particles. Along the bounce-transit orbit, guiding center drifts cause radial excursions from the magnetic field line. In addition, passing particles wind around the magnetic field line toroidally due to the lack of any bounce point. To calculate this frequency, we need to first calculate deviations from the field line orbit, noting that radial excursions from the field line are of the order of the gyroradius. To aid in the calculation, we exploit the exact conservation of the canonical toroidal momentum:
\begin{equation}
\psi = \bar{\psi} + \frac{m B^0_\varphi R_0 v_\parallel} {e B} = \bar{\psi} + \psi_1.
\end{equation}
Here, $\bar{\psi}$ corresponds to the reference magnetic flux surface defined to be
\begin{equation}
\bar{\psi} = - \frac{P_\varphi}{e}
\end{equation} 
and $\psi_1$ is the deviation from that flux surface. Recall that the field line orbit assumed that
\begin{equation}
\frac{d \theta}{dt} \approx \frac{v_\parallel}{q R_0},
\end{equation}
where we hold $r$ and thus $\psi$ fixed. Since the exact field line-following orbit breaks $P_\varphi$ conservation, we need to include deviations from the field line caused by conservation of $P_\varphi$ along with guiding center drifts in order to consistently expand the guiding center equation of motion with respect to the gyroradius.

The guiding center equation of motion is
\begin{equation}
\frac{dx}{dt}  = \left(v_\parallel \hat{b} + \vec{v}_D \right) \cdot \nabla x,
\end{equation}
where $x$ is any spatial coordinate and $\vec{v}_D$ are the guiding center drifts. We then expand the guiding center equation of motion for variables $\psi$, $\theta$, and $\varphi$ and find that
\begin{align}
\frac{d \psi}{dt} & \approx \left. \left(\vec{v}_D \cdot \nabla \psi\right) \right|_{\bar{\psi}}, \\
\frac{d \theta}{dt} & \approx \left. \left(\frac{v_\parallel}{q R_0} \right) \right|_{\bar{\psi}} + \frac{d}{d \psi} \left. \left(\frac{v_\parallel}{q R_0} \right) \right|_{\bar{\psi}} \psi_1 + \left. \left(\vec{v}_D \cdot \nabla \theta \right) \right|_{\bar{\psi}}, \\
\frac{d \varphi}{dt} & \approx \left. \left(\frac{v_\parallel}{q R_0} \frac{d\varphi}{d\theta} \right) \right. + \frac{d}{d \psi} \left. \left(\frac{v_\parallel}{q R_0} \frac{d\varphi}{d\theta} \right)  \right|_{\bar{\psi}} \psi_1 + \left. \left(\vec{v}_D \cdot \nabla \varphi \right) \right|_{\bar{\psi}}.
\end{align}
We take the $\vec{v}_D$ to be the the sum of the classical curvature, grad-$B$, and $E$-cross-$B$ drifts:
\begin{equation}
\vec{v}_D = \frac{\hat{b}}{e B} \left(m v_\parallel^2 \vec{k} + \mu \nabla B + e \nabla \Phi\right),
\end{equation}
where $\hat{k}$ is the curvature vector defined such that
\begin{equation}
\hat{b} \times \vec{k} = \nabla \times \hat{b} - \left(\hat{b} \cdot \nabla \times \hat{b}\right) \hat{b}. 
\end{equation}
We can simplify the equation of motion in the toroidal direction by noting that
\begin{equation}
	\begin{split}
	\frac{d}{d \psi} \left(\frac{v_\parallel}{q R_0} \frac{d\varphi}{d\theta} \right) \psi_1 &=  \frac{d}{d \psi} \left(\frac{v_\parallel}{q R_0} \right) \frac{d\varphi}{d\theta} \psi_1 + \frac{d}{d \psi} \left( \frac{d\varphi}{d\theta}  \right) \frac{v_\parallel}{q R_0} \psi_1 \\ &=  \frac{d\varphi}{d\theta} \frac{d \theta}{dt} + \frac{d}{d \psi} \left( \frac{d\varphi}{d\theta}  \right) \frac{v_\parallel}{q R_0} \psi_1 - \frac{d \varphi}{d\theta} \frac{v_\parallel}{q R_0} - \frac{d \varphi}{d\theta} \left(\vec{v}_D \cdot \nabla \theta \right),
	\end{split}
\end{equation}
where we used the equation of motion in the poloidal direction. Substituting this in and evaluating $\frac{d \varphi}{d\theta}$, we obtain
\begin{equation}
\dot{\varphi} = \frac{d}{d \psi} \left( \frac{q}{1 + \epsilon \cos\left(\theta\right)}\right) \psi_1 \frac{v_\parallel}{q R_0} + \vec{v}_D \cdot \nabla \varphi - \frac{q}{1 + \epsilon \cos\left(\theta\right)} \vec{v}_D \cdot \nabla \theta + \frac{q}{1 + \epsilon \cos\left(\theta\right)} \frac{d \theta}{dt},
\end{equation}
where we evaluate all radial coordinates at $\bar{r}$ such that
\begin{equation}
\psi\left(\bar{r}\right) = \bar{\psi}. 
\end{equation}
Finally, we take the bounce-transit average of $\dot{\varphi}$ and find that 
\begin{equation}
\Omega_3 = \left<\dot{\varphi}\right> = \left<\omega_d\right> + \bar{\epsilon} q(\bar{r})  \Omega_2 = \Omega_d + \bar{\epsilon} q(\bar{r})  \Omega_2.
\end{equation}
Here, $\bar{\epsilon}$ is 0 for trapped particles and 1 for passing particles, $\Omega_d$ is the frequency purely due to the guiding center drifts, and $\omega_d$ is associated with the instantaneous deviation from the magnetic field line. The extra term for passing particles is due to the toroidal rotation from following the field line in a complete poloidal turn. This parallel velocity dependent term is absent for trapped particles since their average toroidal position does not change as a result of a complete field line-following bounce. We approximate $\omega_d$ as
\begin{equation}
\omega_d \approx \frac{d q}{d \psi} \frac{\psi_1 v_\parallel}{q} + \vec{v}_D \cdot \nabla \varphi - q \vec{v}_D \cdot \nabla \theta.
\end{equation}
The poloidal component of the magnetic drift dominates, thus we ignore the toroidal component. Using the $s-\alpha$ equilibrium, we calculate the guiding center drift in Appendix ~\ref{Magnetic}. Thus, $\Omega_3$ is computed with a finite Shafranov shift. The poloidal component of the guiding center drift is
\begin{equation}
\vec{v}_{D} \cdot \nabla \theta \approx -\frac{m}{e B r R_0} \left(v_\parallel^2 + \frac{v_\perp^2}{2} \right) \left( \cos\left(\theta\right) - \alpha \sin^2 \left(\theta\right)\right) - \frac{E_r }{ r B},
\end{equation}
where $E_r$ is the radial electric field. We also define $\alpha$ such that
\begin{equation}
\alpha = - q^2 \beta \frac{R_0}{P} \frac{dP}{dr}.
\end{equation}
Here, $ \beta = 2 \mu_0 P/B^2$ with $\mu_0$ being the vacuum permeability. The $E$-cross-$B$ drift can be separated from the magnetic drifts, so that we obtain 
\begin{equation}
\vec{v}_{D,B} \cdot \nabla \theta \approx - \frac{v_{D,B}}{r} \left(\cos\left(\theta\right)- \alpha \sin^2\left(\theta\right)\right),
\end{equation}
where $v_{D,B}$ characterizes the magnetic drifts and can be written as
\begin{equation}
v_{D,B} = \frac{m}{eBR_0} \left(v_\parallel^2 + \frac{v_\perp^2}{2}\right) = \frac{\xi T}{e B R_0} \left(2 - \lambda b\right),
\end{equation}
where the magnetic shear is
\begin{equation}
	s = \frac{r}{q} \frac{dq}{dr}.
\end{equation}
We can therefore rewrite $ \Omega_d$ as
\begin{equation}
\Omega_d  = \frac{q}{r} \frac{T \xi}{e B R_0} \left<\left(2 - \lambda b\right)\left(\cos(\theta)- \alpha \sin^2(\theta)\right) + \frac{2 s}{\epsilon} \left(1 - \lambda b \right)\right> + \left<\frac{q 	E_r}{r B}\right> = \omega_{d0} \xi F_d(\kappa) + \omega_E, 
\end{equation}
where $F = F_d(\kappa)$ is the bounce-transit averaged term, $\omega_{d0}$ is characteristic of the magnetic precession frequency and defined to be 
\begin{equation}
	\omega_{d0} = \frac{q T}{r e B R_0},
\end{equation}
and $\omega_E$ corresponds to the $E$-cross-$B$ velocity. We explicitly carry out the bounce-transit average by rewriting the $\lambda b$ terms in terms of $\kappa$ and $\theta$ and find that 
\begin{equation}
	F_d(\kappa) \approx \left\{ \begin{aligned}
		 &-1 + \frac{2 E\left(\kappa\right)}{K\left(\kappa\right)} + 4 s \left(\kappa^2 -1 + \frac{E\left(\kappa\right)}{K\left(\kappa\right)} \right) &&\\ & \qquad \qquad - \frac{4 \alpha}{3} \left(1 - \kappa^2 - \left(1 - 2 \kappa^2\right)  \frac{E\left(\kappa\right)}{K\left(\kappa\right)}\right) && \text{ if } 0\le \kappa < 1 \text{ (trapped)},\\
		& - \left(2 \kappa^2 -1 \right)+ 2 \kappa^2 \frac{E\left(\kappa^{-1}\right)}{K\left(\kappa^{-1}\right)} + 4 s \kappa^2 \frac{E\left(\kappa^{-1}\right)}{K\left(\kappa^{-1}\right)} && \\ & \qquad \qquad - \frac{4 \alpha}{3} \kappa^2 \left(\left(2 \kappa^2 -1 \right) \frac{E\left(\kappa^{-1}\right)}{K\left(\kappa^{-1}\right)} - 2 \kappa^2 + 2 \right) && \text{ if } 1 < \kappa < \infty \text{ (passing)}.
	\end{aligned} \right. 
\end{equation}

Rather than calculating $\alpha_3$ explicitly, it suffices to write its generic integral form. We also include the various oscillating quantities associated with the precession motion. In general, we have
\begin{align}
& \psi = \bar{\psi} + \tilde{\psi},\\
& \varphi = \alpha_3 + q(\bar{r}) \tilde{\theta} + \tilde{\varphi},\\
& \theta = \bar{\epsilon} \alpha_2 + \tilde{\theta}. 
\end{align}
Here, $\tilde{\psi}$ represents the excursion from the reference flux surface $\bar{\psi}$ during the poloidal orbit. Meanwhile, $\tilde{\varphi}$ is the difference in toroidal precession between a circular geometry and a more general equilibrium magnetic field. Later, we will use $\tilde{r}$ instead of $\tilde{\psi}$ with the understanding that $\psi\left(\tilde{r}\right) = \tilde{\psi}$. Meanwhile, $\tilde{\theta}$ is associated with the oscillatory poloidal motion. We define these quantities as
\begin{align}
	\tilde{\psi} &= \int^{\alpha_2} \frac{d \alpha_2'}{\Omega_2} \vec{v}_D \cdot \nabla \psi = \psi_1 \\
	\tilde{\theta} &= \int^{\alpha_2} d \alpha_2' \left(\frac{1}{\Omega_2} \frac{d\theta}{dt} - \bar{\epsilon}\right),\\
	\tilde{\varphi} & = \frac{dq}{d\psi} \tilde{\theta} \tilde{\psi} + \int^{\alpha_2}  \frac{d \alpha_2'}{\Omega_2} \left( \bar{\epsilon} \frac{dq}{d \psi} \tilde{\psi} \Omega_2 +   \vec{v}_D \cdot \nabla\varphi - \frac{dq}{d\psi} \tilde{\theta} \vec{v}_D \cdot \nabla \psi - q \vec{v}_D \cdot \nabla \theta - \Omega_d\right).\\
\end{align}
This guarantees that
\begin{equation}
	\frac{d \alpha_3}{dt} = \Omega_d + \bar{\epsilon} q \Omega_2.
\end{equation}
Note that in the above, $q$ and $dq/d\psi$ are evaluated at $\bar{r}$ and thus are time independent. Having characterized action-angle coordinates, we can proceed to solving the Vlasov equation using these coordinates. 

\section{The Vlasov Equation} \label{The Vlasov Equation}
To begin, we write the Vlasov equation in action-angle variables:
\begin{equation}
	\frac{df}{dt} = \frac{\partial f}{\partial t} + \dot{\bm{\alpha}}\cdot \frac{\partial f}{\partial \bm{\alpha}} + \dot{\vec{J}} \cdot \frac{\partial f}{\partial \vec{J}} = 0.
\end{equation}
We remind ourselves that Hamilton's equations of motion in these coordinates are
\begin{align}
	&\dot{\vec{J}} = - \frac{\partial H}{\partial \bm{\alpha}} = - e \frac{\partial \phi}{\partial \bm{\alpha}},\\
	& \dot{\bm{\alpha}} = \frac{\partial H}{\partial \vec{J}} = e \frac{\partial \phi}{\partial \vec{J}} + \bm{\Omega},
\end{align}
We later generalize the above equation with a Krook-style operator to add collisions for trapped electrons in Appendix \ref{Collisions}, but for now we work in the collisionless limit. The next step is to linearize the system by assuming the distribution function is composed of an equilibrium part $f_0 = f_0(\vec{J})$ and a perturbed part $\delta f = \delta f(\bm{\alpha}, \vec{J}, t).$ Dropping any quadratic perturbative terms, we obtain 
\begin{equation}
\frac{\partial \delta f}{\partial t} + \bm{\Omega} \cdot \frac{\partial \delta f}{\partial \bm{\alpha}} - e \frac{\partial \phi}{\partial \bm{\alpha}} \cdot \frac{\partial f_0}{\partial \vec{J}} = 0. 
\end{equation}
As stated earlier, any perturbative functions we consider must be periodic in the angular variables $\bm{\alpha}$. Therefore, we utilize a discrete Fourier transform in $\delta f$ and $\phi$:
\begin{align}
&\delta f = \sum_{\vec{n}} f_{\vec{n}}(\vec{J}) e^{i (\vec{n} \cdot \bm {\alpha} - \omega t)},\\
&\phi = \sum_{\vec{n}} \phi_{\vec{n}}(\vec{J}) e^{i (\vec{n} \cdot \bm {\alpha} - \omega t)}.
\end{align}
 To extract the physical quantity, we take the real part of the Fourier series. Here, $\vec{n}$ corresponds to the mode number of the Fourier term and $\omega$ is the complex frequency of oscillation. We decompose the complex frequency as $\omega = \omega_r + i \gamma$, where $\omega_r$ is the real frequency and $\gamma$ is the growth rate. Note that in QuaLiKiz, we only consider unstable modes with $\gamma > 0$ and ignore stable modes; although this does not change the fundamental approach, it does afford us some slight computational simplicity since we do not have to search for solutions in the entire complex plane. As an ansatz, we treat $\omega = \omega_{n_3}$ to be dependent on $n_3$ only, not $n_1$ and $n_2$. To consistently solve the dispersion relation, we will eventually need to sum over $n_1$ and $n_2$. The individual Fourier components can be calculated from the physical quantity via
\begin{align}
&f_{\vec{n}} = \int \frac{d^3\alpha}{(2 \pi)^3} \delta f(t=0) e^{-i\vec{n} \cdot \bm {\alpha}},\\
&\phi_{\vec{n}} = \int \frac{d^3\alpha}{(2 \pi)^3} \phi(t=0) e^{-i\vec{n} \cdot \bm {\alpha}},
\end{align}
where we integrate each angular variable from $0$ to $2 \pi$.

To proceed, we assume the equilibrium distribution is a shifted Maxwellian:
\begin{equation}
f_0(\vec{J}) = n_0 \left(\frac{m}{2 \pi T}\right)^{3/2} \exp\left(- \frac{m \left(\vec{v} - \vec{U}\right)^2} {2T}\right) = n_0  \left(\frac{m}{2 \pi T}\right)^{3/2} \exp\left( \frac{-H_0 + e \Phi + m \vec{v} \cdot \vec{U} - \frac{m U^2}{2}}{T} \right). 
\end{equation}
Here, $n_0$ is the equilibrium number density, $T$ is the temperature, and $\vec{U}$ is the equilibrium plasma rotation velocity. In general, $n_0$, $T$ and $\vec{U}$ will vary with position and therefore depend on $\vec{J}$. By only considering toroidal rotation, we make the approximation
\begin{equation}
U_\varphi \hat{\varphi} \approx U_\parallel \hat{b},
\end{equation}
which allows us to write
\begin{align}
	U^2 &\approx U_\parallel^2, \\
	\vec{v} \cdot \vec{U} & \approx v_\parallel U_\parallel.
\end{align}
We also take into account gradients of the parallel rotation velocity. Due to the presence of rotation, we also include the radial electric field as well as its gradient. We use the natural natural frequency parameter for the electric field shear $\gamma_E$ defined as
\begin{equation}
	\gamma_E = -\frac{1}{B} \frac{d E_r}{dr}. 
\end{equation}
This will allow us to Taylor expand the characteristic $E$-cross-$B$ frequency such that
\begin{equation}
	\omega_E \approx \omega_{E0} + \omega_E' x, 
\end{equation}
where we expand about the radial distance $x = 0$ and $\partial_r \omega_E = \omega_E'$ is related to the radial electric shear.

Additionally, in QuaLiKiz we ignore terms that go as the square of fluctuating quantities. Since we assume a small Mach number as well as a small derivative in the parallel velocity, we thus assume that
\begin{equation}
	m U_\parallel \left|\frac{\partial v_\parallel}{\partial \vec{J}}\right| \ll \vec{\omega}_\ast. 
\end{equation}
This term is responsible for turbulent acceleration and arises from the presence of rotation in the equilibrium distribution function. We expect this term to be negligible for non-impurities in the low Mach number limit and will thus neglect it as an approximation \citep{garbet2013}.

Substituting the above expressions as well as the Fourier series into the linearized Vlasov equation, we isolate each term mode by mode due to completeness and orthogonality of the Fourier series to find $f_{\vec{n}}$ in terms of $\phi_{\vec{n}}$. The result is
\begin{equation}
f_{\vec{n}} = \frac{e \phi_{\vec{n}} \vec{n} \cdot \frac{\partial f_0}{\partial \vec{J}}}{\vec{n} \cdot \bm{\Omega}  - \omega} = \frac{f_0}{T}\frac{e \phi_{\vec{n}} \vec{n} \cdot \left(\bm{\omega}_\ast + \bm{\omega}_E - \bm{\Omega}\right)}{\vec{n} \cdot \bm{\Omega}  - \omega},
\end{equation}
where the diamagnetic frequency $\bm{\omega}_\ast$ is 
\begin{equation}
\bm{\omega}_\ast = T \left( \frac{1}{n_0} \frac{\partial n_0}{\partial \vec{J}} + \left(\xi - \frac{3}{2} - \frac{U_\parallel}{v_T^2} \left(2 v_\parallel - U_\parallel\right) \right) \frac{1}{T} \frac{\partial T}{\partial \vec{J}}\right) + \frac{2 \left(v_\parallel - U_\parallel \right)}{v_T^2} \frac{\partial U_\parallel}{\partial \vec{J}},
\end{equation}
where the thermal velocity is $v_T = \sqrt{2 T /m }$ and the frequency associated with the $E$-cross-$B$ drift is
\begin{equation}
\bm{\omega}_E = \frac{e}{T} \frac{d \Phi}{d \vec{J}}. 
\end{equation}
We then rewrite the equation to be
\begin{equation}
f_{\vec{n}} = -\frac{e \phi_{\vec{n}}}{T} f_0 \left(1 - \frac{\omega - \vec{n}\cdot \bm{\omega}_\ast - \vec{n} \cdot \vec{\omega}_E}{\omega - \vec{n} \cdot \bm{\Omega}} \right),
\end{equation}
where it is now clear that there is an adiabatic part and a frequency dependent part of the equation. 

The next step to solving the dispersion relation is to use Poisson's equation,
\begin{equation}
\nabla ^2 \phi = \sum_s - \frac{e_s n_s}{\epsilon_0},
\end{equation}
where the $s$ subscript labels the particle species and $\epsilon_0$ is the vacuum permittivity. In the earlier parts of the derivation, we had suppressed the subscript for various quantities (e.g. $m, T, n_0, f_0, \delta f, \ldots$); we include the subscript for the time being. The total number density $n_s$ is
\begin{equation}
n_s = \int d^3 v f_s. 
\end{equation}
The perturbed electrostatic potential is calculated using the perturbed charge density
\begin{equation}
\delta n_s = \int d^3 v \delta f_s.
\end{equation} 
To enforce quasineutrality, we take the sum of the total charge density to be 0 and require that
\begin{equation}
\lambda_D \ll \left|\frac{\phi}{\nabla \phi}\right|,
\end{equation}
where $\lambda_D$ is the Debye length. Because we are interested in length scales much longer than the Debye length, the Laplacian term in Poisson's equation is negligible. We thus obtain
\begin{equation}
\sum_s \int d^3 v e_s \delta f_s = 0. 
\end{equation}
Since $\phi = \phi(\vec{r})$ is independent of velocity, if we multiply both sides of the above equation by $\phi^\ast$, the complex conjugate of $\phi$, we can simply move it inside the integral. We then integrate over space, resulting in
\begin{equation}
\sum_s \int d^3r  d^3 v e_s \phi^\ast \delta f_s = 0. 
\end{equation}
By multiplying by the electrostatic potential and integrating, we have recast the differential equation via a weak formulation using the variational method\citep{saiman1970, garbet1990, garbet2001, nguyen2008}. Instead of solving for the exact function $\phi$ or $\delta f$ that satisfies Poisson's equation, we can simply approximate $\phi$ and $\delta f$ with a suitable function and focus on the dispersion relation itself. Typically, when the Laplacian is kept, the differential equation is put into the weak formulation by integrating the Laplacian term by parts; this technique is well established in other fields such as finite element analysis \citep{johnson1991}.

We next substitute in the Fourier expansions and the expression relating $\phi_{\vec{n}}$ and $f_{\vec{n}}$. The result is
\begin{equation}
\sum_s \sum_{\vec{n}, \vec{n}^\prime} \int d^3r d^3v \frac{e_s^2 \phi_{\vec{n}} \phi_{\vec{n}^\prime}^\ast}{T_s} f_{0,s} \left(1 - \frac{\omega - \vec{n} \cdot \bm{\omega}_\ast - \vec{n} \cdot \bm{\omega}_E}{\omega - \vec{n} \cdot \bm{\Omega}}\right) e^{i (\vec{n} - \vec{n}^\prime) \cdot \bm{\alpha}} e^{-i (\omega - (\omega^\prime)^\ast) t} = 0. 
\end{equation}
To simplify this integral, we first perform the change of variables $(\vec{r}, \vec{v}) \to (\vec{r}, \vec{p})$; the Jacobian of this transformation is simply $m_s^{-1}$. We then perform the change of variables $(\vec{r}, \vec{p}) \to (\bm{\alpha}, \vec{J})$; the Jacobian of this particular transformation is $1$ because this is guaranteed to be a canonical transformation. We therefore obtain
\begin{equation}
\sum_s \sum_{\vec{n}, \vec{n}^\prime} \int d^3\alpha d^3 J \frac{e_s^2 \phi_{\vec{n}} \phi_{\vec{n}^\prime}^\ast}{m_s T_s} f_{0,s} \left(1 - \frac{\omega - \vec{n} \cdot \bm{\omega}_\ast - \vec{n} \cdot \bm{\omega}_E}{\omega - \vec{n} \cdot \bm{\Omega}}\right) e^{i (\vec{n} - \vec{n}^\prime) \cdot \bm{\alpha}} e^{-i (\omega - (\omega^\prime)^\ast) t} = 0. 
\end{equation}
We note that the exponential terms are $\bm{\alpha}$ dependent. We then use orthogonality of the Fourier series to find that
\begin{equation}
\sum_s \sum_{n_1, n_2} \int d^3 J \frac{e_s^2 \left|\phi_{\vec{n}}\right|^2 }{m_s T_s} f_{0,s} \left(1 - \frac{\omega - \vec{n} \cdot \bm{\omega}_\ast - \vec{n} \cdot \bm{\omega}_E}{\omega - \vec{n} \cdot \bm{\Omega}}\right) = 0.
\end{equation}
Because we developed the Fourier series such that $\omega$ only depends on the mode number $n_3$, we can solve for each value of $n_3$ individually while summing over $n_1$ and $n_2$. While the summation arising from this convention seems to make the problem more difficult at first glance, we shall see later it allows for a variety of simplifications. Moreover, the integrand is now completely independent of $\bm{\alpha}$. As such, we integrate over the action angles again and transform back to conventional variables, leading to
\begin{equation}
\sum_s \sum_{n_1, n_2} \int d^3r d^3v \frac{e_s^2 \left|\phi_{\vec{n}}\right|^2 }{T_s} f_{0,s} \left(1 - \frac{\omega - \vec{n} \cdot \bm{\omega}_\ast - \vec{n} \cdot \bm{\omega}_E}{\omega - \vec{n} \cdot \bm{\Omega}}\right) = 0.
\end{equation}
Even though the integrand is a function of only $\vec{J}$, the parameters in the integrand are more naturally expressed in terms of other coordinates such as the minor radius and the pitch angle parameter. Thus, further coordinate transformations to simplify this expression are inevitable. As such, they are most easily carried out when starting from the typical configuration space variables $\left(\vec{r}, \vec{v}\right).$

For ease of notation, we split up the dispersion relation as follows:
\begin{equation}
\sum_s \mathcal{L}_{0, s} - \mathcal{L}_{\text{passing}, s} - \mathcal{L}_{\text{trapped}, s} = 0.
\end{equation}
Here, $\mathcal{L}_0$ is the portion of the integral that is simply multiplied by $1$, which we call the adiabatic part. $\mathcal{L}_{\text{passing}}$ is the portion of the integral that is frequency dependent and integrated over the part of velocity space that encompasses passing particles, while $\mathcal{L}_{\text{trapped}}$ consists of the trapped particles instead.

To proceed with solving the dispersion relation, we must first calculate $\left|\phi_{\vec{n}}\right|^2$. This requires a 3-dimensional integral over $d^3 \alpha$. Once that is done, we then proceed to calculate the integral in the dispersion relation itself for the adiabatic part, trapped part, and passing part separately. Although our expression appears to be a 6-dimensional integral, we can utilize a number of symmetries, transformations, and approximations to simplify the form down to at most 2-dimensional integrals. Although integrals of higher dimension can be in principle calculated numerically, the curse of dimensionality renders such integrals computationally expensive. Thus, a reduction to two dimensions affords us a great deal of speed at the cost of some amount of accuracy.

\section{Ballooning Representation} \label{Ballooning}
Before integrating $\left|\phi_{\vec{n}}\right|^2$ with respect to the action angles, we review key results regarding the ballooning representation. Because $\phi(r,\theta,\varphi)$ must be periodic in $\theta$ and $\varphi$, we may expand $\phi$ as a Fourier series,
\begin{equation}
\phi(r, \theta, \varphi) = \sum_{m, n} \phi_{m,n}(r - r_0) e^{i (m \theta + n \varphi)}. 
\end{equation}
Here, $r_0$ is the location of the resonant flux surface for each given $m$ and $n$; in other words, $q(r_0) = -m/n$. We take these modes to be localized around the resonant flux surface. These modes are often radially localized such that the distance between any two adjacent resonant rational flux surfaces is much longer than the characteristic length scale of the plasma equilibrium. If that condition holds, then all modes $\phi_{n,m}$ all have nearly identical radial envelopes where each radial profile is centered on their corresponding reference flux surface \citep{connor1979}. These flux surfaces are all rational flux surfaces since $m$ and $n$ are integers. Meanwhile, the general ballooning representation of $\phi$ is
\begin{equation}
\phi(r, \theta, \varphi) = \sum_p \sum_n \oldhat{\phi}_{n} (\theta + 2 p \pi, \theta_0 ) e^{in \left(\varphi - q(r) \left(\theta - \theta_0 + 2 p \pi   \right)\right)},
\end{equation}
where $\theta_0$ is the ballooning angle and $p$ denotes the various harmonics. Here, we have approximated the potential by separating it into a quickly varying eikonal and a slowly varying envelope.This representation ultimately comes from the fact that the instabilities in question are strongly anisotropic and flute-like where $k_\parallel \gg k_{\perp}$. In absence of toroidal rotation, the ballooning angle is typically taken to be zero since the most unstable modes are localized around $\theta_b = 0$. In the presence of finite toroidal rotation and an equilibrium electrostatic potential, the ballooning angle is shifted away from zero. However, this shift is typically on the order of $10^{-1}$ in relevant cases \citep{candy2009}. Thus, for the rest of the derivation we take the ballooning angle to be zero as an approximation. This is equivalent to assuming that the envelope is radially independent. Moreover, if the profile is heavily localized around $\theta = 0$, we can use the strong ballooning approximation and ignore all harmonics except for $p = 0$, leading to
\begin{equation}
	\phi(r, \theta, \varphi) = \sum_n \oldhat{\phi}_n(\theta) e^{in \left(\varphi - q(r) \theta\right)}. 
\end{equation}

It is important to note that the decomposition in terms of $\phi_{m,n}$ describes how the same radial profile is localized about adjacent flux surfaces. Meanwhile, the decomposition in terms of $\oldhat{\phi}_n$ describes how the linear eigenmode balloons along the field line. This can be seen more explicitly if one considers that
\begin{align}
	\vec{B} \cdot \nabla \psi &= 0,\\
	\vec{B} \cdot \nabla \left(\varphi - q \theta\right) & = 0,\\
	\vec{B} \cdot \nabla \theta & \neq 0,
\end{align}
indicating that with the above set of variables that $\theta$ indicates the location on any given field line. Because the magnetic curvature is unfavorable on the low field side of the tokamak when one considers the interplay between the curvature vector and the pressure gradient for normal tokamak profiles, we expect fluctuations to peak about $\theta = 0$. We can demonstrate a direct link between $\phi_{m,n}$ and $\oldhat{\phi}_n$ by calculating the Fourier components of $\phi$, leading to 
\begin{equation}
\int_{-\pi}^{\pi} \frac{d \varphi}{2\pi} \int_{-\pi}^{\pi} \frac{d\theta}{2\pi} \phi(r, \theta, \varphi) e^{- i n \varphi - i m \theta} = \phi_{m,n}(r-r_0) = \int_{-\pi}^{\pi} \frac{d \theta}{2 \pi} \oldhat{\phi}_n (\theta) e^{- i \theta (n q(r) + m)}. 
\end{equation}
We then make two approximations. First, we Taylor expand the term in the eikonal around the reference flux surface,
\begin{equation}
n q(r) + m = n q_0 + \frac{r - r_0}{d} + m = \frac{r - r_0}{d} = \frac{x}{d} + \frac{\tilde{r}}{d},
\end{equation}
where $q_0 = q(r_0)$, the radial difference between different rational flux surfaces is defined as,
\begin{equation}
\frac{1}{nd} = \left.\frac{dq}{dr}\right|_{r = r_0},
\end{equation}
and $x$ is defined as	
\begin{equation}
x = \bar{r} - r_0,
\end{equation}
where we ignore second derivatives of the safety factor. After doing so, we find that
\begin{equation}
\phi_{m,n}(r-r_0) \approx \int_{-\pi}^{\pi} \frac{d \theta}{2 \pi} \oldhat{\phi}_n (\theta) e^{- i \frac{\theta \left(r - r_0\right)}{d}}.
\end{equation}
Second, we invoke the strong ballooning approximation by treating $\oldhat{\phi}_n$ as heavily localized around $\theta = 0$; this allows us to integrate from $-\infty$ to $\infty$ instead of from $-\pi$ to $\pi$. The result is
\begin{equation}
\phi_{m,n}(r-r_0) \approx \int_{-\infty}^{\infty} \frac{d \theta}{2 \pi} \oldhat{\phi}_n (\theta) e^{- i \frac{\theta \left(r - r_0 \right)}{d}}. 
\end{equation}
Comparing it with our previous definition of the Fourier transform, we find that $\oldhat{\phi}_n(\theta)$ is simply the Fourier transform of $\phi_{m,n}(r)$, with $k_r = \theta/d$. The transformation is given by
\begin{equation}
\oldhat{\phi}_n(\theta) = \int^{\infty}_{-\infty} \frac{dr}{\left|d\right|} \phi_{m,n}(r-r_0) e^{i \frac{\theta (r-r_0 )}{d}}.
\end{equation}

We are now in a position to integrate over the action angles to fully calculate $\phi_{\vec{n}}$. The procedure to integrate over $\alpha_1$ has already been discussed in Section~\ref{Action Angle Variables}, where we find that
\begin{equation}
	\int_{-\pi}^{\pi} \frac{d \alpha_1}{ 2 \pi} g(\vec{r}) e^{- i n_1 \alpha_1} = \left(-i\right)^{n_{1}} \left(J_{n_{1}} \left(k_\perp \rho\right) \cdot g\right) \left(\vec{R}_G\right). 
\end{equation}
We therefore only need to discuss in detail the integrations over $\alpha_2$ and $\alpha_3$ while treating all variables within the guiding center framework. Trapped particle motion and passing particle motion differ such that the two cases must be handled separately. 

\subsection{Trapped}

For deeply trapped particles, the equations for the action variables simplify to
\begin{align}
r &= \bar{r} + \delta_b \cos(\alpha_2),\\
\theta &= \theta_b \sin(\alpha_2),\\
\varphi &= \alpha_3 + q(\bar{r}) \theta_b \sin(\alpha_2) + \tilde{\varphi}. 
\end{align}
Here, we define the banana width $\delta_b$ as
\begin{equation}
	\delta_b = \frac{q \rho}{\sqrt{\epsilon}}. 
\end{equation}
While more exact expressions for the bounce motion can be given using Jacobi elliptic functions, we use the above equations for all trapped particles as an approximation. We first integrate over $\alpha_2$, once again utilizing the Fourier transform,
\begin{equation}
\int^{\pi}_{-\pi} \frac{d \alpha_2}{2 \pi} \phi(\vec{r}) e^{- i n_2 \alpha_2} = \int^{\pi}_{-\pi} \frac{d \alpha_2}{2 \pi} \int \frac{d^3 k}{(2 \pi)^3}\tilde{\phi}(\vec{k}) e^{-i \vec{k} \cdot \vec{r} - i n_2 \alpha_2}.
\end{equation}
We then proceed in fashion similar to the gyro-average derivation in Section~\ref{Action Angle Variables} by noting that $\vec{k} \cdot {\vec{r}} = \vec{k} \cdot \bar{\vec{r}} + k_r \delta_b \cos(\alpha_2)$. The result is
\begin{equation}
\int^{\pi}_{-\pi} \frac{d \alpha_2}{2 \pi} \phi(\vec{r}) e^{- i n_2 \alpha_2} = \int \frac{d^3 k}{(2 \pi)^3} (-i)^{n_2} J_{n_2}(k_r \delta_b) \tilde{\phi}(\vec{k}) e^{-i \vec{k} \cdot \bar{\vec{r}}}.
\end{equation}
In essence, we obtain a bounce average over the banana width \citep{depret2000}. We note that this in particular is a rather crude approximation. The particularities of the bounce motion such as the bounce angle and the radial excursion technically depend on the pitch angle of the particle; we are in essence smearing this out by taking a representative trapped particle such that the banana width is constant. The averaging procedure is also approximate as we only take into account the radial deviation. As seen in \citet{bilgari1986}, we would normally obtain a $\theta$ dependence in the argument of the Bessel function; the $k_r$ term manifests as the Fourier link established earlier. As a result, trapped particles have two Bessel operators acting on the potential corresponding to the gyromotion and the banana orbit respectively.

We now proceed to integrating over $\alpha_3$, for now leaving the Bessel functions aside and evaluating the position at $\vec{r} = \bar{\vec{r}}$. In doing so, we must be aware that for trapped particles $\bar{\theta} = 0$; that is, the variation of $\theta$ only comes from the bounce orbit which we averaged over. We also ignore $\tilde{\varphi}$ for the same reason. Moreover, because we assume the modes to have an identical radial structure, we are free to keep only one of the poloidal harmonics. Making the strong assumption that the actual radial envelope can be approximated in this way, we pick $m_0 = - n_3 q_0$, as this forces any $\theta$ dependence we approximately neglected in the eikonal to vanish. Thus, we obtain
\begin{equation}
\int_{-\pi}^{\pi} \frac{d \alpha_3}{2\pi} \phi(\bar{\vec{r}}) e^{- i n_3 \alpha_3} = \sum_{n} \int_{-\pi}^{\pi} \frac{d \alpha_3}{2\pi} \phi_{m_0,n}(\bar{r}- r_0) e^{i \left(n \alpha_3 - n_3 \alpha_3\right)} = \phi_{m_0,n_{3}}(\bar{r}- r_0). 
\end{equation}
To compensate for choosing only one poloidal harmonic, we must extend the radial limits of integration to $-\infty < r < \infty$. Aside from the Bessel functions, nothing in the trapped part of the dispersion relation is dependent on $\theta = k_r d$. Therefore, we are free to take the amplitude squared of the averaged potential to obtain
\begin{equation}
\left|\phi_{\vec{n}}\right|^2 = \left|J_{n_1}(k_\perp \rho)J_{n_2}(\delta_b k_r) \cdot \phi_{m_0, n_3}\right|^2(\bar{r} - r_0) = \left|J_{n_1}(k_\perp \rho)J_{n_2}(\delta_b k_r) \cdot \phi_{m_0, n_3}\right|^2(x),
\end{equation}
where we evaluate the function at $x = \bar{r} - r_0$.

\subsection{Passing}
We now calculate $\phi_{\vec{n}}$ for passing particles. Instead of utilizing the poloidal harmonics, it is more useful to use the ballooning representation directly. Substituting in the expression for $\alpha_3$ and then integrating over $\alpha_3$ leads to
\begin{equation}
\int_{-\pi}^{\pi}\frac{d \alpha_3}{2 \pi} \phi(\vec{r}) e^{-i n_3 \alpha_3} = \sum_n \int_{-\pi}^{\pi}\frac{d \alpha_3}{2 \pi} \oldhat{\phi}_n(\theta(\alpha_2))e^{i n(\alpha_3 + \tilde{\varphi} - q(r) \alpha_2 +(q(\bar{r}) - q(r)) \tilde{\theta}) - i n_3 \alpha_3}.
\end{equation}
Here, $\theta$ is taken to be a function of $\alpha_2$. It is crucial that we recognize not all the safety factors in the eikonal are evaluated at the same point. We have both $q(r) = q(\bar{r} + \tilde{r})$ and $q(\bar{r})$. The term $q(r) - q(\bar{r})$ can be Taylor expanded about $r_0$:
\begin{equation}
q(r) - q(\bar{r}) \approx q_0 + \frac{r - r_0}{nd} - q_0 - \frac{\bar{r}-r_0}{nd} = \frac{r - \bar{r}}{nd} = \frac{\tilde{r}}{nd} .
\end{equation} 
Carrying out the integral then gives us
\begin{equation}
\int_{-\pi}^{\pi}\frac{d \alpha_3}{2 \pi} \phi(\vec{r}) e^{-i n_3 \alpha_3} = \oldhat{\phi}_{n_3}(\theta(\alpha_2)) e^{i n_3 (\tilde{\varphi} - q(r) \alpha_2 - \frac{\tilde{r}}{n_3 d} \tilde{\theta})}.
\end{equation}
We now multiply by $e^{- i n_2 \alpha_2}$ and integrate with respect to $\alpha_2$. The eikonal can be simplified if we only keep $n_2 = m_0 = - n q_0$,
\begin{equation}
i \left(n_3 \tilde{\varphi} - n_3 q(r) \alpha_2 - \frac{\tilde{r}}{d} \tilde{\theta} - m_0 \alpha_2\right) = i \left(n_3 \tilde{\varphi} - \frac{x}{d} \alpha_2 - \frac{\tilde{r}}{d}\left(\alpha_2 + \tilde{\theta})\right)\right) = i \left(n_3 \tilde{\varphi} - \frac{x}{d} \alpha_2 - \frac{\tilde{r}}{d} \theta(\alpha_2)\right),
\end{equation}
where we have used
\begin{equation}
-n_3 q(r) \alpha_2 - m_0 \alpha_2 \approx -n_3 q_0 \alpha_2 - \frac{r - r_0}{d} \alpha_2 - m_0 \alpha_2 = -\frac{r - r_0}{d} \alpha_2 = -\frac{x}{d} \alpha_2 - \frac{\tilde{r}}{d} \alpha_2
\end{equation}
and the expression $\alpha_2 = \theta - \tilde{\theta}$. We then obtain
\begin{equation}
\int_{-\pi}^{\pi} \frac{d \alpha_2}{2 \pi}  \oldhat{\phi}_{n_3} (\theta(\alpha_2))e^{i \left(n_3 \tilde{\varphi} - \frac{x}{d} \alpha_2 - \frac{\tilde{r}}{d} \theta(\alpha_2)\right)} \approx
\int_{-\infty}^{\infty} \frac{d \alpha_2}{2 \pi}  \oldhat{\phi}_{n_3} (\theta(\alpha_2))e^{i \left(n_3 \tilde{\varphi} - \frac{\tilde{r}}{d} \theta(\alpha_2)\right)} e^{-i\frac{x}{d}\alpha_2},
\end{equation}
where we have invoked the strong ballooning approximation. We can see that this is simply an inverse Fourier transform going from $\alpha_2$ to $x$. Thus, we write that
\begin{equation}
\phi_{\vec{n}} = \phi_{n_1, m_0, n_3} = \left(J_{n_1}(k_\perp \rho) \cdot \mathcal{F}^{-1}\left(\oldhat{\phi}_{n_3} (\theta(\alpha_2))e^{i \left(n_3 \tilde{\varphi} - \frac{\tilde{r}}{d} \theta(\alpha_2)\right)}\right)\right) (x),
\end{equation}
where $\mathcal{F}^{-1}$ inverts the Fourier transform as described above with respect to $\alpha_2$. While the $\tilde{\varphi}$ dependence can be approximately ignored in a circular geometry, the $\tilde{r} \theta(\alpha_2)$ dependence in the eikonal must be kept. 

 We shall see that the mode numbers $n_1$ and $n_2$ do not appear explicitly in the final expression. For convenience, we thus write $n_3 = n$ and identify it as the toroidal mode number.

\subsection{Gaussian Eigenfunction}

We now introduce the functional form of the potential. We use the ansatz that the poloidal harmonic structure is a shifted Gaussian: 
\begin{equation}
	\phi_{m_0, n}(x) \sim \phi_0 e^{-\frac{\left(x - x_0\right)^2}{2 w^2}}. 
\end{equation}
This Gaussian corresponds to the lowest-order eigenfunction from the corresponding ballooning equation; the higher-order terms utilize the Hermite polynomials and are neglected here. This Gaussian has a complex width $w$ and shift $x_0$. In the limit of no rotation, $x_0 = 0$ and the Gaussian is centered about $x = 0$. Although the amplitude $\phi_0$ cannot be obtained from quasilinear theory, it factors out of the dispersion relation and does not affect the linear mode frequency calculation. Setting the amplitude will be necessary to calculate the quasilinear fluxes and requires the use of a saturation rule, which is detailed in Section~\ref{Saturation}. 

To obtain expressions for $w$ and $x_0$, we move into the high-frequency fluid limit. The original derivation can be found in \citet{cottier2014} and an extensive, revised derivation can be found in \citet{citrin2017}; here, we shall only discuss the basic principle. We consider the dispersion relation
\begin{equation}
	D\left(\omega\right) = \sum_s \int d^3 v \frac{f_0 e_s^2}{T_s} \left(1 - J_{0,s}^2 \frac{\omega -n \omega_E - n \omega_{\ast, s}}{\omega - n \omega_E - k_\parallel v_\parallel - n \omega_{d,s}}\right) \phi_{m_0, n}(x). 
\end{equation}
This is the local dispersion relation obtained if we consider the strong form of Poisson's equation rather than the weak form; we do not multiply by $\phi^\ast$ and integrate over space. The Bessel function is such that $J_{0,s} = J_0\left(k_\perp \rho_s\right) J_0\left(k_r \delta_{b,s}\right)$ for trapped particles and $J_{0,s} = J_0\left(k_\perp \rho_s\right)$ for passing particles. Meanwhile, we define the parallel wave number as $k_\parallel = \left(k_\theta s x\right)/\left(q R_0\right)$. The local drift frequency $\omega_{d,s}$ is
\begin{equation}
	\omega_{d,s} = \left\{ \begin{aligned} 	
	& \omega_{d0, s} \xi \Biggl( -1 + \frac{2 E\left(\kappa\right)}{K\left(\kappa\right)} + 4 s \left(\kappa^2 -1 + \frac{E\left(\kappa\right)}{K\left(\kappa\right)} \right) &&\\ & \qquad \qquad \quad - \frac{4 \alpha}{3} \left(1 - \kappa^2 - \left(1 - 2 \kappa^2\right)  \frac{E\left(\kappa\right)}{K\left(\kappa\right)}\right)\Biggr) && \text{ if trapped},\\
	& \omega_{d0, s} \xi \left(2 - \lambda b \right) \left( \cos\left(\theta\right) + \left(s \theta - \alpha \sin\left(\theta\right)\right) \sin\left(\theta\right) \right) && \text{ if passing}.  
	\end{aligned} \right. 
\end{equation}
The passing form of the drift frequency is due to the radial structure of the eigenfunction as covered in Section~\ref{Passing}. We also note the Fourier link in the passing drift frequency that $\theta^2 \to k_r^2 d^2$. To proceed, we take $\overbar{\omega} = \omega - n \omega_E$ to be larger than $k_\parallel v_\parallel$ and $\omega_{d0,s}$, and for trapped particles we approximate $k_\parallel v_\parallel \approx 0$. We also take $\delta_{b,e} \ll \delta_{b,i}$ and $\rho_e \ll \rho_i$, where the ``e'' subscript is for electrons and the ``i'' subscript is for ions, to obtain
\begin{equation}
	\begin{split}
	D\left(\omega\right) = \Biggl[\frac{n_e}{T_e} &\left(1 - \left<\left(1 - \frac{n \omega_{\ast,e}}{\overbar{\omega}}\right) \left(1 + \frac{n \omega_{d,e}}{\overbar{\omega}} + \frac{n^2 \omega_{d,e}^2}{\overbar{\omega}^2} \right)    \right>_t \right) \\ &+ \sum_i \frac{n_i Z_i}{T_i} \Biggl(1 - \left< \left(1 - \frac{n \omega_{\ast,i}}{\overbar{\omega}} \right) \left(1 + \frac{n \omega_{d,i}}{\overbar{\omega}} + \frac{n^2 \omega_{d,i}^2}{\overbar{\omega}^2} \right) \left(1 - \frac{k_r^2 \delta_{b,i}^2}{2} \right) \left(1 - \frac{k_\theta^2 \rho_i^2}{2}\right) \right>_t  \\ & \qquad - \left< \left(1 - \frac{n \omega_{\ast,i}}{\overbar{\omega}} \right) \left(1 + \frac{n \omega_{d,i}}{\overbar{\omega}} + \frac{k_\parallel v_\parallel}{\overbar{\omega}} 	+ \left(\frac{n \omega_{d,i}}{\overbar{\omega}} + \frac{k_\parallel v_\parallel}{\overbar{\omega}} \right)^2 \right) \right. \\ & \qquad \qquad \times \left. \left(1 - \frac{k_\theta^2 \rho_i^2}{2} - \frac{k_r^2 \rho_i^2}{2} \right) \right>_p \Biggr)
	\Biggr] \phi_{m_0, n}(x), 
	\end{split}
\end{equation}
where $Z_i$ is the proton number of the ion species. We define the averages over velocity space as
\begin{align}
	\left<g\left(\vec{v}\right)\right>_t & = \int_{\text{trapped}} d^3v f_0 g\left(\vec{v}\right) , \\
	\left<g\left(\vec{v}\right)\right>_p & = \int_{\text{passing}} d^3v f_0 g\left(\vec{v}\right). 
\end{align}
Do to the $\theta$ dependent terms, this is a differential equation. We approximate the differential operators on $\phi$ in the limit of small mode shift $x_0$, leading to
\begin{equation}
	\frac{\partial^2 \phi}{\partial x^2} = \left(\frac{x^2}{w^4} - \frac{2 x_0 x}{w^4} - \frac{1}{w^2} \right) \phi. 
\end{equation}

Next, we carry out the integrals both analytically and numerically as appropriate and multiply the dispersion relation by $\omega^3$ to obtain a modified dispersion relation,
\begin{equation}
	\omega^3 D\left(\omega\right) = D_0 \left(\omega\right) + D_1 \left(\omega\right) x + D_2 \left(\omega\right) x^2 = 0.	
\end{equation}
Here, we separate terms proportional to $x^0, x^1$, and $x^2$. With three equations we can solve for the three unknowns $(\omega_0, w, x_0)$. We then find the solution $\omega_0$ such that
\begin{equation}
	D_0 \left(\omega_0 \right) = 0. 
\end{equation}
Having found this zeroth-order solution, we then find $x$ and $w$ such that
\begin{align}
	D_1 \left(\omega_0\right) = 0,\\
	D_2 \left(\omega_0 \right) = 0. 
\end{align}
We do not cite the full solution here and direct the reader to \citet{citrin2017} for a complete derivation. Now that we have characterized $\phi_{\vec{n}}$ by calculating $x_0$ and $w$, we move to the dispersion relation itself, beginning with the adiabatic term.

\section{Adiabatic Functional} \label{Adiabatic}

We first examine the adiabatic part of the functional, as it is the simplest to treat. It takes the form
\begin{equation}
\mathcal{L}_0 = \sum_{n_1, n_2} \int d^3r d^3v \frac{e^2 \left|\phi_{\vec{n}}\right|^2}{ T} f_0.
\end{equation}
Here, we have suppressed the subscript $s$ as we will be working with each species independently. We first define a new function $\phi_{n} = \phi_{n}(\alpha_1, \alpha_2, \vec{J})$ such that 
\begin{equation}
\phi_{n} = \int_{-\pi}^{\pi} \frac{d \alpha_3}{2\pi} \phi(\vec{r}) e^{-i n \alpha_3} = \sum_{n_1, n_2} \phi_{\vec{n}} e^{i (n_1 \alpha_1 +n_2 \alpha_2 )}.
\end{equation}
We then note due to the orthogonality of the Fourier series that
\begin{equation}
\int_{-\pi}^{\pi} \frac{d \alpha_1}{2 \pi} \int_{-\pi}^{\pi} \frac{d\alpha_2}{2\pi} \left|\phi_{n}\right|^2 = \sum_{n_1, n_2} \left|\phi_{\vec{n}}\right|^2.
\end{equation}
Thus, it is more convenient to switch back to action-angle coordinates for an intermediate calculation: 
\begin{equation}
\mathcal{L}_0 = \sum_{n_1, n_2} \int d^3\alpha d^3J \frac{e^2 \left|\phi_{\vec{n}}\right|^2}{m T} f_0 = \sum_{n_1, n_2} \int 4 \pi^2 d \alpha_3 d^3J \frac{e^2 \left|\phi_{\vec{n}}\right|^2}{m T} f_0 = \int 4 \pi^2 d \alpha_3 d^3J \frac{e^2 \left|\phi_n\right|^2}{m T} f_0. 
\end{equation}
This then simplifies to
\begin{align}
\mathcal{L}_0 = \int 4 \pi^2 d\alpha_3 d^3J \frac{e^2}{m T} f_0 \int_{-\pi}^{\pi} \frac{d \alpha_1}{2 \pi} \int_{-\pi}^{\pi} \frac{d\alpha_2}{2\pi} \left|\phi_{n}\right|^2 = \int d^3r d^3v \frac{e^2 \left|\phi_{n}\right|^2}{T}  f_0. 
\end{align}

The velocity space integration is straightforward,
\begin{equation}
\mathcal{L}_0 = \int d^3r d^3v \frac{e^2 \left|\phi_{n}\right|^2}{T}  f_0 = \int d^3r \frac{e^2 n_0 \left|\phi_{n}\right|^2}{T},
\end{equation}
so all that is left is the spatial integration. Because we use toroidal coordinates, the differential volume element is
\begin{equation}
d^3r = r R_0 \left(1 + \epsilon \cos\left(\theta\right)\right) dr d\theta d\varphi.
\end{equation} 

We proceed to calculating $\phi_{n}$ using the poloidal harmonic expansion as detailed in Section \ref{Ballooning},
\begin{equation}
\phi(r, \theta, \varphi) = \sum_{m, n} \phi_{m,n}(r - r_0) e^{i (m \theta + n \varphi)}. 
\end{equation}
When we examined the trapped Fourier modes, we already calculated $\phi_{n}$. We simply need to generalize it for passing particles as well, resulting in
\begin{equation}
\phi_{n} = \sum_m \phi_{m, n}(r - r_0) e^{i (m \theta + n q(\bar{r}) \tilde{\theta} + n \tilde{\varphi})}.
\end{equation}
As before, we only keep the poloidal harmonic corresponding to $m_0 = -n q_0$ and expand the limits of integration for $r$ to compensate. The result is
\begin{equation}
\left|\phi_{n}\right|^2 = \left|\phi_{m_0, n}(r-r_0)\right|^2.  
\end{equation}
Because the integrand in the adiabatic functional now only depends on $r$, the integral simplifies to
\begin{equation}
\mathcal{L}_0 = \int_{-\infty}^{\infty} \left(2 \pi R_0\right) 2\pi r dr \frac{e^2 n_0}{T} \left|\phi_{m_0, n}(r-r_0)\right|^2 \approx \int dx  R_0 r_0 \left(2 \pi\right)^2 \frac{e^2 n_0}{T} \left|\phi_{m_0, n}(x)\right|^2.
\end{equation}
Here, we make use of the localization approximation which transforms the factor of $r$ in the integrand into $r_0$. Due to the Gaussian structure of $\phi_{m_0, n}$, this integral is easily performed and we find that
\begin{equation}
\mathcal{L}_0 = 4 \pi^2 R_0 r_0 \left|\phi_0\right|^2 \left|w\right|^2  \exp\left(\frac{\Imag\left(x_0\right)^2} {\Real\left(w^2\right)}\right) \sqrt{\frac{\pi}{\Real\left(w^2\right)}}. 
\end{equation}
Now that we have calculated the adiabatic functional, we next calculate the trapped functional. 

\section{Trapped Functional} \label{Trapped}
The trapped part of the dispersion relation reads
\begin{equation}
\mathcal{L}_{\text{trapped}} = \sum_{n_1, n_2} \int d^3r d^3v \frac{e^2}{T} f_0 \left(\frac{\vec{n} \cdot \bm{\omega}_\ast + \vec{n} \cdot \vec{\omega}_E - \omega}{\vec{n} \cdot \bm{\Omega} - \omega}\right) |\phi_{\vec{n}}|^2. 
\end{equation}
We emphasize that although this aspect of the derivation is collisionless, QuaLiKiz includes collisions for trapped electrons. Strictly speaking, this section concerns trapped ions. The majority of the derivation remains the same for trapped electrons, the key difference being that the eventual integral over the particle energy cannot be analytically simplified. 

The first step is to determine the appropriate variables to integrate over. For the spatial variables, we use once again use toroidal coordinates, 
\begin{equation}
	d^3r = R_0 \bar{r} \left(1 + \epsilon \cos\left(\theta\right)\right) d\bar{r} d\theta d\varphi. 
\end{equation}
For velocity space, we use the variables $\left(v, \lambda, v_\phi\right)$ which correspond to the speed $v$, pitch angle parameter $\lambda$, and cylindrical velocity phase $v_\phi$. The result is
\begin{equation}
	d^3 v = \sum_{\epsilon_\parallel} v^2 \frac{b}{2\sqrt{1 - \lambda b}} dv d\lambda dv_\phi,
\end{equation}
where the sum over $\epsilon_\parallel$ accounts for both possible signs of the parallel velocity. Because the integrand is independent of $\varphi$ or $v_\phi$, we obtain
\begin{equation}
	d^3r d^3v = \sum_{\epsilon_\parallel} 2 \pi^2 R_0 \bar{r} \left(1 + \epsilon \cos\left(\theta\right)\right) d\bar{r} d\theta  v^2 \frac{b}{\sqrt{1 - \lambda b}} dv d\lambda.
\end{equation}

It is important to note that the limits of integration depend on the order of integration. For a given $\theta$, the pitch angle parameter $\lambda$ for a trapped particle is bounded by
\begin{equation}
	\frac{1 - \epsilon}{1 + \epsilon} \le \lambda \le \frac{1 + \epsilon \cos\left(\theta\right)}{1 + \epsilon}.
\end{equation} 
 The lower bound corresponds to the trapped-passing boundary, while the upper bound corresponds to a particle that has $v_\parallel = 0$ at a given angle $\theta$. We can, however, exchange the order of integration as follows:
\begin{equation}
	\int_{0}^{2 \pi} d \theta \int_{\frac{1-\epsilon}{1+\epsilon}}^{\frac{1+ \epsilon \cos\left(\theta\right)}{1 + \epsilon}} d\lambda f(\theta, \lambda) = \int_{\frac{1-\epsilon}{1+\epsilon}}^{1} d\lambda \int_{-\theta_b}^{\theta_b} d\theta f(\theta, \lambda). 
\end{equation}
We recall that the definition of a bounce average is
\begin{equation}
\left<G(\epsilon_\parallel, \theta)\right> = \frac{\int_{-\theta_b} ^{\theta_b} d\theta \frac{G\left(\epsilon_\parallel, \theta\right) + G\left(-\epsilon_\parallel, \theta\right)}{\sqrt{1-\lambda b}} } { \int_{-\theta_b} ^{\theta_b}\frac{2 d\theta}{\sqrt{1-\lambda b}} } = \sum_{\epsilon_\parallel} \frac{\overbar{\Omega}_2}{2 \pi} \int_{-\theta_b} ^{\theta_b} d\theta \frac{G\left(\epsilon_\parallel, \theta\right)}{\sqrt{1-\lambda b}}. 
\end{equation}
By exchanging our limits of integration and integrating over $\theta$ first, part of the trapped functional simplifies to become a bounce average. 

Next, we approximate the equilibrium distribution function assuming the Mach number $U_\parallel ~/ ~c_s$ is small, where $c_s = \sqrt{T/m}$ is the sound speed. Since the electron and ion rotation velocity is quite small compared to the sound speed in tokamak plasmas, expanding to second-order in the Mach number will be sufficient. The distribution function then simplifies to
\begin{equation}
f_0 \approx n_0 \left(\frac{m}{2 \pi T}\right)^{3/2} e^{-\xi} \left( 1 + \frac{2 v_\parallel U_\parallel}{v_T^2} + \frac{U_\parallel^2}{v_T^2} \left(\frac{2 v_\parallel^2}{v_T^2} - 1\right) \right).  
\end{equation}
Moreover, because $\left|\Omega_1\right|, \left|\Omega_2\right| \gg \left|\omega\right|$, we can approximate this integral by truncating the sum at $n_1 = n_2 = 0$. We also perform a change of variables from $v$ to $\xi$ to obtain 
\begin{equation}
	\begin{split}
	\mathcal{L}_{\text{trapped}} = \int d \bar{r} & d \lambda d \xi \left(2 \pi\right)^2 \frac{n_0 e^2}{T} \frac{R_0 \bar{r}}{\sqrt{\pi}} \frac{\sqrt{\xi} e^{-\xi}}{\overbar{\Omega}_2} \\ & \times \frac{\left< \left(\vec{n} \cdot \vec{\omega}_\ast + \vec{n} \cdot \vec{\omega}_E - \omega \right) \left( 1 + \frac{2 v_\parallel U_\parallel}{v_T^2} + \frac{U_\parallel^2}{v_T^2} \left(\frac{2 v_\parallel^2}{v_T^2} - 1\right) \right) \right>}{n \Omega_3 - \omega} \left|\phi_{0,0,n}\right|^2. 
	\end{split}
\end{equation}
The Bessel functions from the gyromotion and the banana motion  are implicit in $\phi_{0,0,n}$. We next simplify the partial derivatives with respect to $\vec{J}$. Because $n_1 = n_2 = 0$, we only keep the partial derivative with respect to $J_3$. Knowing that $J_3 = J_3\left(\bar{r}\right)$, we perform a change in variables from $J_3$ to $\bar{r}$ and find that
\begin{equation}
	\frac{\partial g}{\partial  J_3} = \frac{\partial g}{\partial \bar{r}} \frac{d \bar{r}}{d J_3} \approx \frac{-R_0 \omega_{d0}}{T} \frac{\partial g}{\partial \bar{r}},
\end{equation}
where $g$ is a generic scalar function. We then define the following normalized gradients:
\begin{align}
	A_n &= - \frac{R_0}{n} \frac{d n}{d\bar{r}},\\
	A_T &= - \frac{R_0}{T} \frac{d T}{d \bar{r}}, \\
	A_U &= - \frac{R_0}{v_T} \frac{d U_{\parallel}}{d \bar{r}}.
\end{align}
To perform the bounce average, we note that only $v_\parallel$ is dependent on $\theta$. We perform the calculation explicitly to find that
\begin{align}
	\left<v_\parallel\right> &= \left<v_\parallel^3\right> =0 ,\\
	\left<v_\parallel^2\right> & = \frac{4 E \epsilon}{m} \frac{\left(E\left(\kappa\right) - \left(1 - \kappa^2\right) K\left(\kappa\right)\right)}{K\left(\kappa\right)} = v_T^2 \xi H\left(\kappa\right),
\end{align}
where we define
\begin{equation}
	H\left(\kappa\right)= \frac{2 \epsilon  \left(E\left(\kappa\right) - \left(1 - \kappa^2\right) K\left(\kappa\right)\right)}{K\left(\kappa\right)}. 
\end{equation}
To simplify our expressions, we also fold $\vec{n} \cdot \vec{\omega}_E$ into the mode frequency such that
\begin{equation}
	\overbar{\omega} = \omega - \vec{n} \cdot \vec{\omega}_E \approx \omega - n \omega_E \approx \omega - n \omega_{E0} - n \omega_E' x, 
\end{equation}
where, as discussed earlier, we Taylor expand $\omega_E$ about $x = 0$ and $\partial_r \omega_E = \omega_E'$ is related to the radial electric shear. Rather than including $x$ fully, we instead approximate the term by averaging it over the Gaussian eigenfunctions: 
\begin{equation}
	\left<x\right>_r = \frac{\int_{-\infty}^{\infty} dx \left|\phi_{m_0, n}\left(x\right)\right|^2 x }{\int_{-\infty}^{\infty} dx \left|\phi_{m_0, n}\left(x\right)\right|^2} = \Real\left(x_0\right) +  \frac{\Imag\left(x_0\right) \Imag\left(w^2\right)}{\Real\left(w^2\right)}. 
\end{equation}
We then obtain
\begin{equation}
	\overbar{\omega} \approx \omega - n \omega_{E0} - n \omega_E' \left<x\right>_r. 
\end{equation}

Ignoring all terms that are order cubic or higher with the Mach number, we then find that
\begin{equation}
	\left< \left( \vec{n} \cdot \vec{\omega}_\ast + \vec{n} \cdot \vec{\omega}_E - \omega \right) \left(1 + \frac{2 v_\parallel U_\parallel}{v_T^2} + \frac{U_\parallel^2}{v_T^2} \left( \frac{2 v_\parallel^2}{v_T^2} - 1\right)\right) \right > = n\omega_{d0}\left( \mathcal{A}_t + \mathcal{B}_t \xi + \mathcal{C}_t \xi^2 	\right),
\end{equation}
where
\begin{align}
	\mathcal{A}_t & = \left(1 - \frac{U_\parallel^2}{v_T^2}\right) \left(A_n - \frac{3}{2} A_T - z^2 F_d \left(\kappa\right) \right) - \frac{U_\parallel}{v_T} \left(2 A_U - \frac{U_\parallel}{v_T} A_T\right), \\
	\mathcal{B}_t & = \left(1 - \frac{U_\parallel^2}{v_T^2}\right) A_T + 4 A_U \frac{U_\parallel}{v_T} H\left(\kappa\right) + \frac{U_\parallel^2}{v_T^2} H\left(\kappa\right) \left(2 A_n - 7 A_T - 2 z^2 F_d \left(\kappa\right) \right) \nonumber \\ & \approx \left(1 - \frac{U_\parallel^2}{v_T^2}\right) A_T + 4 A_U \frac{U_\parallel}{v_T} H\left(\kappa\right), \\
	\mathcal{C}_t & = 2 A_T \frac{U_\parallel^2}{v_T^2} H\left(\kappa\right) \approx 0.
\end{align}
Here, we have defined
\begin{equation}
	z^2 = \frac{\overbar{\omega}}{n \omega_{d0} F_d \left(\kappa\right)}. 
\end{equation}
Moreover, we take note that $H\left(\kappa\right) \sim \mathcal{O}\left(\epsilon\right)$; since the inverse aspect ratio $\epsilon$ is small, we can safely ignore all terms proportional to $U_\parallel^2 H\left(\kappa\right) / v_T^2$. 

Substituting the above into the integrand, we obtain
\begin{equation}
	\mathcal{L}_\text{trapped} = \int d \bar{r} d \lambda d \xi \left(2 \pi\right)^2 \frac{n_0 e^2}{T} \frac{R_0 \bar{r}}{\sqrt{\pi}} \frac{\sqrt{\xi} e^{-\xi}}{\overbar{\Omega}_2} \left|\phi_{0,0,n}\right|^2 \frac{\mathcal{A}_t + \mathcal{B}_t \xi} {F_d \left(\kappa\right)\left(\xi - z^2\right)}. 
\end{equation}
Due to the localization of the mode, we evaluate any functions of $\bar{r}$ at $r_0$ in the above expression aside from the electrostatic potential. We then rewrite the trapped functional as
\begin{equation}
	\mathcal{L}_\text{trapped} = \int d \bar{r} d \lambda d \xi \left(2 \pi\right)^2 \frac{n_0 e^2}{T} \frac{R_0 r_0}{\sqrt{\pi}} \frac{\sqrt{\xi} e^{-\xi}}{\overbar{\Omega}_2} \left|J_0\left(k_\perp \rho\right) J_0 \left(k_r \delta_b\right) \cdot \phi_{m_0, n}\right|^2 \frac{\mathcal{A}_t + \mathcal{B}_t \xi} {F_d \left(\kappa\right)\left(\xi - z^2\right)}. 
\end{equation}
The gyromotion and bounce motion appear in two separate Bessel functions. Since the only explicit radial dependence is contained in the electrostatic potential, we can change variables using Parseval's theorem to integrate over $k_r$,
\begin{equation}
	\int_{-\infty}^{\infty} dx f(x) g(x)^\ast = \int_{-\infty}^{\infty} \frac{d k_r}{2\pi} \oldhat{f}(k_r) \oldhat{g}(k_r)^\ast. 
\end{equation}
After transforming to Fourier space, we treat the Bessel functions as normal scalar functions instead of differential operators. We next note that the Bessel functions are dependent on velocity through the gyroradius and banana width,
\begin{align}
	\rho &= \frac{v_\perp}{\Omega_1},\\
	\delta_b & \approx \frac{q}{\sqrt{\epsilon}} \rho.
\end{align}
We approximate this energy dependence by averaging each Bessel function separately over velocity space using a Maxwellian distribution. Doing so allows us to retain finite Larmor radius and finite banana width effects while also making the energy and pitch angle integration tractable. We find that
\begin{equation}
\frac{\int d^3v J_0\left(k_\perp \rho\right)^2 f_0}{\int d^3v f_0} = e^{-\frac{k_\perp^2 \rho_{\text{th}}^2}{2}} I_0\left(\frac{k_\perp^2 \rho_{\text{th}}^2}{2}\right) = \Gamma_0 \left(k_\perp \rho_{\text{th}}\right),
\end{equation}
where $I_0$ is a modified Bessel function of the first kind and the characteristic thermal gyroradius $\rho_\text{th}$ is defined as
\begin{equation}
\rho_{\text{th}} = \frac{\sqrt{2T/m}}{\Omega_1}. 
\end{equation}
Similarly, for the average over the banana orbit we obtain
\begin{equation}
\frac{\int d^3v J_0\left(k_\perp \rho\right)^2 f_0}{\int d^3v f_0} = e^{-\frac{k_\perp^2 \delta_{b,\text{th}}^2}{2}} I_0\left(\frac{k_\perp^2 \delta_{b,\text{th}}^2}{2}\right) = \Gamma_0 \left(k_\perp \delta_{b,\text{th}}\right),
\end{equation}
where  the thermal banana width is
\begin{equation}
\delta_{b, \text{th}} = \frac{q}{\sqrt{\epsilon}} \rho_\text{th}. 
\end{equation}
Note that $k_\perp^2$ is written as 
\begin{equation}
	k_\perp^2 = k_r^2 + k_\theta^2 =  k_r^2 + \frac{n^2 q_0^2}{r_0^2},
\end{equation}
where we have evaluate $k_\theta$ at $r_0$. That $k_\theta^2 = n^2q^2/r^2$ comes from differentiating with respect to $\theta$ in the ballooning expansion due to the eikonal term. Because the $k_r$ dependence is now completely separable from the $\kappa$ and $\xi$ dependence, we write the trapped functional as 
\begin{equation}
	\begin{split}
	\mathcal{L}_\text{trapped} = \int d \lambda d \xi \left(2 \pi\right)^2 &\frac{n_0 e^2}{T} \frac{R_0 r_0}{\sqrt{\pi}} \frac{\sqrt{\xi} e^{-\xi}}{\overbar{\Omega}_2}  \frac{\mathcal{A}_t + \mathcal{B}_t \xi} {F_d \left(\kappa\right)\left(\xi - z^2\right)} \\ & \times \int_{-\infty}^{\infty} \frac{dk_r}{2 \pi} \Gamma_0 \left(k_\perp \rho_\text{th}\right) \Gamma_0 \left(k_r \delta_{b,\text{th}}\right) \left| d\oldhat{\phi}_n \left(k_r d\right) \right|^2,
	\end{split}
\end{equation}
where $\oldhat{\phi}_n$ is computed using a Fourier transform: 
\begin{equation}
	\oldhat{\phi}_n \left(k_r d\right) = \sqrt{2 \pi} w \phi_0 e^{-\frac{k_r^2 w^2}{2}} e^{i k_r x_0}. 
\end{equation}

We next simplify the integral over $\xi$, which is of the form
\begin{equation}
	\int_{0}^{\infty} d \xi \frac{\sqrt{\xi}}{\sqrt{\pi}} \frac{\mathcal{A}_t + \mathcal{B} \xi}{\xi - z^2} e^{-\xi} = \int_{-\infty}^{\infty} du \frac{u^2}{\sqrt{\pi}} \frac{\mathcal{A}_t + \mathcal{B}_t u^2}{\left(u + z\right)\left(u - z\right)} e^{-u^2},
\end{equation}
where we performed the change of variables $\xi = u^2$. Using the plasma dispersion function detailed in Appendix~\ref{Fried and Conte}, this simplifies to
\begin{equation}
	\int_{-\infty}^{\infty} du \frac{u^2}{\sqrt{\pi}} \frac{\mathcal{A}_t + \mathcal{B}_t u^2}{\left(u+z\right)\left(u-z\right)} e^{-u^2} = \mathcal{A}_t G_2\left(z, -z\right) + \mathcal{B}_t G_4\left(z, -z\right) = \frac{\mathcal{A}_t Z_2 \left(z\right) + \mathcal{B}_t Z_4 \left(z\right)}{z},
\end{equation}
where the final simplification is made using the fact that $Z_{2n}$ is an even function for $n \ge 0$. Meanwhile, we rewrite the integration over $\lambda$ with a change in variables,
\begin{equation}
	\frac{d \lambda}{\overbar{\Omega}_2} = 4 f_t K\left(\kappa\right) \kappa d \kappa, 
\end{equation}
where we utilize the transformation
\begin{equation}
	\lambda \approx 1 - 2 \epsilon \kappa^2	
\end{equation}
and define the flux surface averaged trapped particle fraction
\begin{equation}
	f_t = \frac{2 \sqrt{2 \epsilon}}{\pi}. 
\end{equation}
Thus, the trapped functional simplifies to
\begin{equation}
	\mathcal{L}_\text{trapped} = \left(2 \pi\right)^3 \frac{e^2 n_0}{T} r_0 R_0 f_t \left<\mathcal{I}_t\right>_{\xi, \kappa} \left<\Gamma_0 \left(k_\perp \rho_{\text{th}}\right) \Gamma_0 \left(k_r \delta_{b, \text{th}}\right) \left|d \oldhat{\phi}\left(k_r d\right)\right|^2\right>_{k_r}, 
\end{equation}
where
\begin{equation}
	\left<\mathcal{I}_t\right>_{\xi, \kappa} = \frac{2}{\pi} \int_0^1 d \kappa \frac{K\left(\kappa\right) \kappa}{z F_d \left(\kappa\right)} \left(\mathcal{A}_t Z_2\left(z\right) + \mathcal{B}_t Z_4 \left(z\right)\right)
\end{equation}
and
\begin{equation}
	\left<\Gamma_0 \left(k_\perp \rho_{\text{th}}\right) \Gamma_0 \left(k_r \delta_{b, \text{th}}\right) \left|d \oldhat{\phi}\left(k_r d\right)\right|^2\right>_{k_r} = \int_{-\infty}^{\infty} \frac{d k_r}{2 \pi} \Gamma_0 \left(k_\perp \rho_{\text{th}}\right) \Gamma_0 \left(k_r \delta_{b, \text{th}}\right) \left|d \oldhat{\phi}\left(k_r d\right)\right|^2. 
\end{equation}
The remaining integrals are to be calculated numerically, where we note that $z$ is a function of both $\overbar{\omega}$ and $\kappa$. Thus, the trapped functional is the product of two separate $1$-dimensional integrals, one of which is $\omega$ independent; we therefore characterize the trapped functional as a $1$-dimensional integral that must be calculated numerically. Now that we have simplified the expression for the trapped functional, we turn to calculating the passing functional.

\section{Passing Functional} \label{Passing}
The passing part of the dispersion relation reads
\begin{equation}
	\mathcal{L}_{\text{passing}} = \sum_{n_1, n_2} \int d^3r d^3v \frac{e^2}{T} f_0 \left(\frac{\vec{n} \cdot \bm{\omega}_\ast + \vec{n} \cdot \vec{\omega}_E - \omega}{\vec{n} \cdot \bm{\Omega} - \omega}\right) |\phi_{\vec{n}}|^2. 
\end{equation}
We reuse many of the same arguments in Section~\ref{Trapped} regarding changes in variables and approximating the equilibrium distribution function. One key difference is that instead of the bounce average, we use the transit average
\begin{equation}
	\left<G\left(\epsilon_\parallel, \theta\right) \right> = 
	\frac{\int_{-\pi}^{\pi} d \theta \frac{G\left(\epsilon_\parallel, \theta\right)}{\sqrt{1 - \lambda b}}} {\int_{-\pi}^{\pi} \frac{ d \theta}{\sqrt{1 - \lambda b}}}
		= \frac{\overbar{\Omega}_2}{2 \pi} \int_{-\pi}^{\pi} d\theta \frac{G\left(\epsilon_\parallel, \theta\right)}{\sqrt{1 - \lambda b}}. 
\end{equation}
We note here that the bounce angle $\theta_b$ is set to $\pi$ and that we do not perform a sum over $\epsilon_\parallel$ to compute the transit average. Moreover, the integration bounds for $\lambda$ are such that
\begin{equation}
	0 \le \lambda \le \frac{1 - \epsilon}{1+\epsilon}.
\end{equation}
These bounds hold regardless of whether we integrate over $\theta$ before or after integrating over $\lambda$. Since they are independent of $\theta$, the order of integration of the two variables can be freely interchanged. As in the trapped case, we only keep $n_1 = 0$ since $\left|\Omega_1\right| \gg \left|\omega\right|$. As discussed in Section~\ref{Ballooning}, $n_2$ refers to the poloidal harmonic. We keep only $n_2 = m_0$ and use the approximation that
\begin{equation}
	m_0 + n q\left(\bar{r}\right) \approx \frac{x}{d}. 
\end{equation}
In the resonant denominator we then obtain
\begin{equation}
	n \Omega_3 + m_0 \Omega_2 - \omega \approx n \omega_{d0} \xi F + \frac{x}{d} \Omega_2 + n \omega_{E0} + n \omega_{E}' x - \omega,
\end{equation}
where we also expand $\omega_E$ about $x = 0$. The passing functional is then
\begin{equation}
	\begin{split}
	\mathcal{L}_{\text{passing}} = \sum_{\epsilon_\parallel} \int d \bar{r} & d \lambda d \xi \left(2 \pi\right)^2 \frac{n_0 e^2}{T} \frac{R_0 \bar{r}}{\sqrt{\pi}} \frac{\sqrt{\xi} e^{-\xi}}{\overbar{\Omega}_2} \\ & \times \frac{\left< \left(\vec{n} \cdot \vec{\omega}_\ast + n \omega_{E0} + n \omega_E' x - \omega \right) \left( 1 + \frac{2 v_\parallel U_\parallel}{v_T^2} + \frac{U_\parallel^2}{v_T^2} \left(\frac{2 v_\parallel^2}{v_T^2} - 1\right) \right) \right>}{n \omega_{d0} \xi F + \frac{x}{d} \Omega_2 + n \omega_{E0} + n \omega_{E}' x - \omega } \left|\phi_{0,m_0,n}\right|^2. 
	\end{split}
\end{equation}
Here, we have evaluated all functions at $r = r_0$ except for the terms proportional to $x$ in the resonant denominator and numerator. These terms must be kept if we wish to take into account the effects of the poloidal motion as well as the radial electric field shear. We now evaluate the integration over $\bar{r}$ while leaving aside the term proportional to $x$ in the numerator.  

To proceed, we use Parseval's theorem to integrate over $\alpha_2$ instead of $\bar{r}$,
\begin{equation}
\int_{-\infty}^{\infty} d\bar{r} f(\bar{r}) g(\bar{r})^\ast = \int_{-\infty}^{\infty} \frac{d \alpha_2}{2\pi} \left|d\right| \oldhat{f}(\alpha_2) \oldhat{g}(\alpha_2)^\ast. 
\end{equation}
For convenience, we compute the radial integral in isolation and relabel variables,
\begin{equation}
	\int_{-\infty}^{\infty} d \bar{r} \frac{1}{a \frac{x}{d} - b} \left|\phi_{0, m_0, n}\right|^2 = \int_{-\infty}^{\infty} \frac{d \alpha_2}{2 \pi} \left|d\right| \mathcal{F}\left(\frac{\phi_{0, m_0, n} }{a \frac{x}{d} - b} \right) \mathcal{F}\left( \phi_{0, m_0, n}\right)^\ast. 
\end{equation}
We calculated in Section~\ref{Ballooning} that
\begin{equation}
	\mathcal{F}\left(\phi_{0, m_0, n}\right) = J_0(k_\perp(\alpha_2) \rho) \oldhat{\phi}_n (\theta(\alpha_2)) e^{i n \tilde{\varphi}(\alpha_2) - i \frac{\tilde{r}(\alpha_2)}{d} \theta(\alpha_2)}.
\end{equation}
We note that $k_\perp$ is defined such that
\begin{equation}
	k_\perp\left(\alpha_2\right)^2 = \frac{\theta\left(\alpha_2\right)^2}{d^2} + \frac{n^2 q_0^2}{r_0^2}.
\end{equation}
We next use the convolution theorem to calculate the other Fourier transform,
\begin{equation}
	\mathcal{F} \left( \frac{\phi_{0, m_0, n} }{a \frac{x}{d} - b}\right) = \frac{1}{2 \pi} \mathcal{F}\left( \phi_{0, m_0, n} \right) \ast \mathcal{F}\left( \frac{1 }{a \frac{x}{d} - b}\right). 
\end{equation}
Computing the Fourier transform of both functions and performing the convolution, we find that
\begin{equation}
	\begin{split}
	\mathcal{F} \left( \phi_{0, m_0, n} \frac{1}{1 + a \frac{x}{d} - b}\right) = \int_{-\infty}^{\infty} &d \alpha_2' \frac{i}{\left|a\right|} \Theta\left(\frac{\alpha_2 - \alpha_2'}{a}\right)  J_0(k_\perp(\alpha_2') \rho) \oldhat{\phi}_n (\theta(\alpha_2')) \\ & \times e^{\frac{-ib}{a} \left(\alpha_2 ' - \alpha_2\right)} e^{i n \tilde{\varphi}(\alpha_2') - i \frac{\tilde{r}(\alpha_2)}{d} \theta(\alpha_2')},
	\end{split}
\end{equation}
where $\Theta$ is the Heaviside step function. Here, we have assumed that $\Imag\left(b\right) > 0$. This is justified since $b \sim \omega$ and we are only interested in positive growth rates. Combining the results, we obtain for the passing integral $I_{p,r}$ that
\begin{equation}
	\begin{split}
		I_{p,r} = \int d\bar{r} \frac{1}{a \frac{x}{d} - b } \left|\phi_{0,m_0,n}\right|^2 =
		\int \frac{d \alpha_2 d \alpha_2'}{2\pi} \frac{i |d|} {\left|a\right|}& \Theta\left(\frac{\alpha_2 - \alpha_2'}{a}\right) J_0^\ast (k_\perp(\alpha_2) \rho) J_0(k_\perp(\alpha_2') \rho)\\ & \times  \oldhat{\phi}_n(\theta(\alpha_2))^\ast \oldhat{\phi}_n(\theta(\alpha_2'))  e^{\Lambda(\alpha_2) - \Lambda(\alpha_2')},
	\end{split}
\end{equation}
where 
\begin{equation}
	\Lambda(\alpha_2) = i \left( \frac{b}{a} \alpha_2 - n \tilde{\varphi}\left(\alpha_2\right) + \frac{\tilde{r}\left(\alpha_2\right)}{d} \theta\left(\alpha_2\right) \right). 
\end{equation}
We then substitute in
\begin{align}
	a &= \Omega_2 + n d \omega_E', \\
	b &= \omega - n \omega_{d0} \xi F - n \omega_{E0}
\end{align}
and rewrite the eikonal term to obtain
\begin{equation}
	\Lambda(\alpha_2) = i \left( \frac{\omega - n \omega_{d0} \xi F - n \omega_{E0}}{\Omega_2 + n d \omega_E'} \alpha_2 - n \tilde{\varphi}\left(\alpha_2\right) + \frac{\tilde{r}\left(\alpha_2\right)}{d} \theta\left(\alpha_2\right) \right). 
\end{equation}
It is important to recognize the physical importance of $\Lambda$. In the ballooning representation, we encoded a certain particle trajectory in the eikonal that differs from the magnetic drift trajectory. The function $\Lambda$ encapsulates the phase difference between these two trajectories. 

Before proceeding, we must recognize that integrating over $\alpha_2$ and $\alpha_2'$ is inconvenient. The function $\oldhat{\phi}_n$ has Gaussian structure in $\theta$, but not in $\alpha_2$. Thus, the next goal is to write the integrand in terms of $\theta$ and $\theta'$. First, we introduce new variables,
\begin{align}
	\theta_+ &= \frac{\theta + \theta'}{2},\\
	\theta_- &= \theta - \theta'.
\end{align}
We next Taylor expand the terms in the exponential around $\theta_+$ to find that
\begin{equation}
	\Lambda(\alpha_2) - \Lambda(\alpha_2') \approx \theta_- \frac{d \alpha_2}{d\theta}\left(\theta = \theta_+ \right) \Lambda^\prime \left(\theta = \theta_+ \right),
\end{equation}
where $\Lambda'$ denotes the derivative of $\Lambda$ with respect to $\alpha_2$. Due to the rapidly varying phase in the exponential, the factor of $i \omega \sim -\gamma$ in the exponential, and the Gaussian integrand, we ignore higher order terms to obtain the dominant contribution. From the equations listed in Section \ref{Action Angle Variables}, we find
\begin{equation}
	\frac{d}{d \alpha_2} \left(n \tilde{\varphi} - \frac{\tilde{r}}{d} \theta\right) =  \frac{n}{\Omega_2} \left(\vec{v}_D \cdot \nabla \varphi - q \vec{v}_D \cdot \nabla \theta - \theta \frac{dq}{d\psi} \vec{v}_D \cdot \nabla \psi - \Omega_d \right). 
\end{equation}
The leading terms can be computed explicitly in much the same manner as when calculating the magnetic precession frequency,
\begin{equation}
	\begin{split}
	\overbar{\Omega}_d &= \vec{v}_D \cdot \nabla \varphi - q \vec{v}_D \cdot \nabla \theta - \theta \frac{dq}{d\psi} \vec{v}_D \cdot \nabla \psi \\ &= \omega_{d0} \xi \left(2 - \lambda b \right) \left(\cos\left(\theta\right) + \left(s \theta - \alpha \sin\left(\theta\right) \sin^2\left(\theta\right) \right)\right) + \frac{q E_r}{r B}.
	\end{split}
\end{equation}
Although somewhat similar to the magnetic drift frequency proper, there are two key differences. Firstly, the magnetic shear term is different and proportional to $\theta \sin\left(\theta\right)$. Secondly, this frequency is explicitly $\theta$ dependent and no bounce-transit average is performed. In carrying out the calculation the bounce-averaged magnetic drift terms partially cancel; for sufficiently small radial electric field shear we obtain
\begin{equation}
	\Lambda\left(\alpha_2\right) - \Lambda\left(\alpha_2'\right) \approx -i \theta_- \left(\frac{n \overbar{\Omega}_d \left(\theta_+ \right) - \omega}{\Omega_2 + n d \omega_E'}\right),
\end{equation}
where
\begin{equation}
	n \overbar{\Omega}_d \approx n \omega_{d0} \xi \left(2 - \lambda b \right) \left(\cos\left(\theta\right) + \left(s \theta - \alpha \sin\left(\theta\right) \sin^2\left(\theta\right) \right)\right) + n \omega_{E0}.
\end{equation}
We next change the variables of integration from $\alpha_2, \alpha_2'$ to $\theta, \theta'$ using
\begin{equation}
	\frac{d \theta}{d \alpha_2} = \frac{\sqrt{1 - \lambda b}}{\overbar{\Omega}_2}. 
\end{equation}
The integral then becomes
\begin{equation}
	\begin{split}
	I_{p,r} = \int \frac{d \theta d \theta'}{2\pi} & \frac{i \left|d\right|}{\left|\Omega_2 + n d \omega_E' \right|}  \Theta\left(\frac{\theta_-}{\Omega_2 + n d \omega_E' }\right)  J_0 \oldhat{\phi}_n(\theta)^\ast J_0' \oldhat{\phi}_n(\theta') \\ & \times  \frac{\overbar{\Omega}_2}{\sqrt{1 - \lambda b(\theta)}} \frac{\overbar{\Omega}_2}{\sqrt{1 - \lambda b(\theta')}} e^{-i \theta_- \frac{\overbar{\Omega}_2}{\sqrt{1 - \lambda b(\theta_+)}} \left(\frac{n \overbar{\Omega}_d (\theta_+) - \omega}{\Omega_2 + n d \omega_E'}\right)},
	\end{split}
\end{equation}
where the Bessel functions are evaluated in terms of $\theta$ and $\theta'$. 
We now substitute in an expression for $\oldhat{\phi}_n$ in terms of a Fourier transform to obtain
\begin{equation}
	\begin{split}
	I_{p,r} = \int \frac{d \theta d \theta' dx dx'}{2\pi } & \frac{i \overbar{\Omega}_2^2}{\left|\Omega_2 + n d \omega_E' \right| \left|d\right|}  \Theta\left(\frac{\theta_-}{\Omega_2 + n d \omega_E'}\right)   J_0 \phi^\ast_{m_0,n}(x) J_0' \phi_{m_0,n}(x') \\ & \times \frac{e^{-i \theta_- \frac{\overbar{\Omega}_2}{\sqrt{1 - \lambda b(\theta_+)}} \left(\frac{n \overbar{\Omega}_d(\theta_+) - \omega}{\Omega_2 + n d \omega_E'}\right)} e^{- i \frac{\theta x}{d}} e^{i \frac{\theta' x'}{d}}}{\sqrt{1 - \lambda b\left(\theta\right)} \sqrt{1 - \lambda b\left(\theta'\right)}}. 
	\end{split}
\end{equation}
We then make the following substitutions
\begin{align}
	x_+ &= \frac{x + x'}{2},\\
	x_- &= x - x',\\
	k_+ &= \frac{\theta_+}{\left|d\right|},\\
	k_- &= \frac{\theta_-}{\left|d\right|},\\
	d\theta d\theta' dx dx' &= dk_+ dk_- dx_+ dx_- \left|d\right|^2,\\
	\frac{\theta'x'}{d} - \frac{\theta x}{d} & = \left(- k_- x_+ - k_+ x_- \right) \frac{\left|d\right|}{d},
\end{align}
to obtain
\begin{equation}
	\begin{split}
	I_{p,r} = 	\int  \frac{d k_+ dk_- dx_+ dx_-}{2\pi } & \frac{i  \left|d\right| \left(\overbar{\Omega}_2\right)^2}{|\Omega_2 + n d \omega_E'|}  \Theta\left(\frac{k_-}{\Omega_2 + n d \omega_E'}\right) J_0 \phi^\ast_{m_0,n}\left(x_+ + \frac{x_-}{2}\right) J_0' \phi_{m_0,n}\left(x_+ - \frac{x_-}{2}\right)\\ & \times \frac{ \exp \left( \frac{-i k_- \left|d\right|\overbar{\Omega}_2}{\sqrt{1 - \lambda b(k_+ \left|d\right|)}} \left(\frac{n \overbar{\Omega}_d(k_+ \left|d\right|) - \omega}{\Omega_2 + n d \omega_E'}\right)\right) e^{-i \left(k_- x_+ + k_+ x_- \right) \frac{\left|d\right|}{d}}}{\sqrt{1 - \lambda b\left(k_+ \left|d\right| + \frac{k_- \left|d\right| }{2} \right)}\sqrt{1 - \lambda b\left(k_+ \left|d\right| - \frac{k_- \left|d\right| }{2} \right)}}  . 
	\end{split}
\end{equation}

At first glance, it seems like we have only made the derivation more difficult. We are now performing a 4-dimensional integration over variables which do not have a convenient Gaussian structure. Fortunately, this simplifies. First, we notice that the integration over $k_-$ via an integration by parts procedure. In general, for a complex parameter $c$ we obtain
\begin{equation}
	\begin{split}
	\int_{a}^{b} ds g \left(s\right) e^{i c s} &= \int_{a}^{b} ds \frac{1}{ik} \left(\frac{d}{ds} \left(g\left(s\right) e^{i c s}\right) - \frac{dg}{ds} e^{i c s}\right) \\ & = \left. \frac{g\left(s\right) e^{i c s}}{i c} \right|_{a}^{b} - \int_a^b ds \frac{1}{\left(i c \right)^2} \left( \frac{d}{ds} \left(\frac{dg}{ds} e^{i c s}\right) - \frac{d^2 g}{ds^2} e^{i cs}\right). 
	\\ & \approx \sum_{m=0}^{N} \frac{\left(-1\right)^m}{\left( i c\right)^{m+1}}\left[ \frac{d^m g}{ds^m} \left(b\right) - \frac{d^m g}{ds^m} \left(a\right)\right]. 
	\end{split}
\end{equation}
This is the asymptotic expansion for sufficiently large $c$. We apply a similar expansion to the integral over $k_-$ and keep only the first term. Because $\Imag\left(\omega\right) > 0$ and the integrand contains a Heaviside step function, the first term is guaranteed to converge. Note that we would normally need to apply the method of steepest descent to properly approximate the integral; however, this requires that the term in the exponential have a saddle point somewhere in the complex plane. Due to our previous approximation, the term in the exponential is monotonic in $k_-$, thus the method of steepest descent is not necessary for our purposes. We find then that
\begin{equation}
	\begin{split}
	\int_{-\infty}^{\infty} dk_- \frac{i \left|d\right|}{ \left|\Omega_2 + n d \omega_E'\right|} & g\left( k_- \right)  e^{- i k_- \frac{\left|d\right|}{d}} \exp\left( \frac{-i k_- \left|d\right|\overbar{\Omega}_2}{\sqrt{1 - \lambda b(k_+ \left|d\right|)}} \left(\frac{n \overbar{\Omega}_d(k_+ \left|d\right|) - \omega}{\Omega_2 + n d \omega_E'}\right)\right) \\ & \approx \frac{g(0)}{\frac{\overbar{\Omega}_2}{\sqrt{1 - \lambda b \left(k_+ \left|d\right| \right)}} \left(n \overbar{\Omega}_d \left(k_+ \left|d\right| \right) - \omega\right) + \Omega_2 \frac{x_+}{d} + n \omega_E' x_+}. 
	\end{split}
\end{equation}	
For convenience, we next replace all instances of $k_+ \left| d \right|$ with $k_+ d$; this is allowed since $\overbar{\Omega}_d$ and $b$ are even functions and the bounds of integration are symmetric. We obtain 
\begin{equation}
	I_{p,r} \approx \int \frac{d k_+ dx_+ dx_-}{ 2 \pi} \frac{\overbar{\Omega}_2  e^{- i k_+ x_-}}{\sqrt{1 - \lambda b \left(k_+ d\right)}}  \frac{J_0\left(\rho k_\perp\right)^2 \phi_{m0,n}\left(x_+ + \frac{x_-}{2}\right)^\ast \phi_{m0,n}\left(x_+ - \frac{x_-}{2}\right) }{n \overbar{\Omega}_d \left(k_+ d\right) + \left(\frac{\Omega_2}{\overbar{\Omega}_2} \frac{x_+}{d} + n \frac{\omega_E'}{\overbar{\Omega}_2} x_+\right)\sqrt{1 - \lambda b\left(k_+ d\right)} - \omega},
\end{equation}	
where
\begin{equation}
	k_\perp^2 = k_+^2 + \frac{n^2 q_0^2}{r_0^2}. 
\end{equation}
As with the trapped functional, we separately average over the Bessel functions,
\begin{equation}
	\frac{\int d^3v J_0 \left(k_\perp \rho\right)^2 f_0}{\int d^3v f_0} = \Gamma_0 \left(k_\perp \rho_\text{th}\right). 
\end{equation}
We next carry out the integral over $x_-$ by identifying it as the inverse Fourier transform of the product of two Gaussians, leading to
\begin{equation}
	\begin{split}
	I_{p,r} = \int \frac{dx_+ dk_+}{\sqrt{\pi}} \frac{\overbar{\Omega}_2}{\sqrt{1 - \lambda b \left(k_+ d\right)}}\frac{ \Gamma_0 \left(k_\perp \rho_{\text{th}}\right) \sqrt{\frac{\Imag\left(w^2\right)^2}{\Real\left(w^2\right)} + \Real\left(w^2\right)} e^{-\rho_\ast^2} e^{-k_\ast^2} \exp\left(\frac{\Imag\left(x_0\right)^2}{\Real\left(w^2\right)}\right) }{n \overbar{\Omega}_d \left(k_+ d\right) + \left(\frac{\Omega_2}{\overbar{\Omega}_2} \frac{x_+}{d} + n \frac{\omega_E'}{\overbar{\Omega}_2} x_+\right)\sqrt{1 - \lambda b\left(k_+ d\right)} - \omega}, 
	\end{split}
\end{equation}
where 
\begin{align}
	\rho_\ast & = \frac{x_+ + k_+ \Imag\left(w^2\right) - \Real\left(x_0\right)}{\sqrt{\Real\left(w^2\right)}},\\
	k_\ast & = \frac{k_+ \Real\left(w^2\right) + \Imag \left(x_0\right)}{{\sqrt{\Real\left(w^2\right)}}}.
\end{align}
It is more convenient to numerically integrate this over $k_\ast$ and $\rho_\ast$ to take advantage of the explicit Gaussian structure. Because the Jacobian of this variable transformation is $1$, the change of variables is easily carried out. In addition, we approximate the $\lambda$ dependent terms by averaging over the pitch angle parameter. We also use the extremely-passing particle limit, where $\theta \approx \alpha_2$. We then obtain
\begin{equation}
	\begin{split}
		I_{p,r} \approx 
		\int_{-\infty}^{\infty} \int_{-\infty}^{\infty} \frac{d\rho_\ast dk_\ast}{\sqrt{\pi}} \frac{ \Gamma_0 \left(k_\perp \rho_{\text{th}}\right) \sqrt{\frac{\Imag\left(w^2\right)^2}{\Real\left(w^2\right)} + \Real\left(w^2\right)} e^{-\rho_\ast^2} e^{-k_\ast^2} \exp\left(\frac{\Imag\left(x_0\right)^2}{\Real\left(w^2\right)}\right) }{n \omega_{d0} \xi F_p\left(k_+ d\right) + \epsilon_\parallel \sqrt{\xi} \frac{x_+}{d}   \frac{\sqrt{2T/m}}{q R_0} - \overbar{\omega}},
	\end{split}
\end{equation}
where
\begin{equation}
	\overbar{\omega} = \omega - n \omega_{E0} - n \omega_E' x_+,
\end{equation}
and
\begin{equation}
	F_p\left(k_+ d\right) = \frac{4}{3} \left(\cos\left(k_+ d\right) + \left(s k_+ d - \alpha \sin\left(k_+ d\right)\right)\sin\left(k_+ d\right)\right).
\end{equation}
We note here that the factor of $4/3$ comes from taking the pitch angle average of $2 - \lambda b$ in the small $\epsilon$ limit. This approximation can be improved by considering higher-order $\epsilon$ terms, although this is not done in the current formulation of QuaLiKiz. 

We now address terms in the numerator of the original integrand that are proportional to $x$; these terms arise from the radial electric field shear. In principle, their inclusion can be treated fully consistently by using the appropriate Fourier transforms as well as the convolution theorem in much the same way we did before. However, as a crude approximation, we simply map $x \to x_+$ in the numerator as is effectively done in the denominator. 

Next, we address the integration over $\lambda$ and $\xi$ in the full passing functional. Once these integrals are calculated, we fold them into the integration over $\rho_\ast$ and $k_\ast$. We wish to compute 
\begin{equation}
	I_{p,E} = \sum_{\epsilon_\parallel} \int d \lambda d \xi \left(2 \pi\right)^2 \frac{n_0 e^2}{T} \frac{R_0 \bar{r}}{\sqrt{\pi}} \frac{\sqrt{\xi} e^{-\xi}}{\overbar{\Omega}_2} \frac{\left< \left(\vec{n} \cdot \vec{\omega}_\ast - \overbar{\omega} \right) \left( 1 + \frac{2 v_\parallel U_\parallel}{v_T^2} + \frac{U_\parallel^2}{v_T^2} \left(\frac{2 v_\parallel^2}{v_T^2} - 1\right) \right) \right>}{n \omega_{d0} \xi F_p\left(k_+ d\right) + \epsilon_\parallel \sqrt{\xi} \frac{x_+}{d}   \frac{\sqrt{2T/m}}{q R_0} - \overbar{\omega}}.
\end{equation}
Because we averaged out the pitch angle dependence in the denominator of the integrand, the pitch angle integration in the numerator is completely separable and only dependent on the inverse aspect ratio $\epsilon$. This is perhaps the largest single approximation used in the passing part of the dispersion; it is necessary to ensure that the numerical integral is 2-dimensional rather than 3-dimensional. It is of potential interest to study the impact this approximation has; one could calculate a more exact (albeit slower) integral to quantify the exact impact this has on the resulting solutions and flux calculations.

Since only $v_\parallel$ terms in the numerator are dependent on $\lambda$. We also use the fact that
\begin{equation}
	\int_{0}^{\frac{1-\epsilon}{1+\epsilon}} \frac{d \lambda }{\overbar{2 \Omega}_2} = f_p, 
\end{equation}
where $f_p = 1 - f_t$ is the flux surface averaged passing particle fraction. We then compute
\begin{equation}
	\int_{0}^{\frac{1-\epsilon}{1+\epsilon}} \frac{d \lambda}{\overbar{\Omega}_2} \left<v_\parallel^m\right>  = 2 f_p v_T^{m} \epsilon_\parallel^m \xi^{m/2} \lambda_m,
\end{equation}
 where we define $\lambda_m$ as
\begin{equation}
	\lambda_m = \frac{\int_{0}^{\frac{1-\epsilon}{1+\epsilon}}d\lambda \int_{-\pi}^{\pi} \frac{d \theta \left(\sqrt{1 - \lambda b}\right)^m}{\sqrt{1-\lambda b}}}{\int_{0}^{\frac{1-\epsilon}{1+\epsilon}}d\lambda \int_{-\pi}^{\pi} \frac{d \theta}{\sqrt{1-\lambda b}}}.
\end{equation}
We numerically calculate $\lambda_m$ separately from the rest of the dispersion relation since $\lambda_m$ is only dependent on $\epsilon$. Once again ignoring terms order cubic or higher with the Mach number, we find that
\begin{equation}
	\begin{split}
	\int_0^{\frac{1-\epsilon}{1+\epsilon}}  \frac{d\lambda }{\overbar{\Omega}_2} &\left< \left( \vec{n} \cdot \vec{\omega}_\ast - \overbar{\omega} \right) \left(1 + \frac{2 v_\parallel U_\parallel}{v_T^2} + \frac{U_\parallel^2}{v_T^2} \left( \frac{2 v_\parallel^2}{v_T^2} - 1\right)\right) \right > = \\ & \qquad \qquad 2 f_p n\omega_{d0}\left( \mathcal{A}_p + \mathcal{B}_p  \epsilon_\parallel \xi^{1/2}  +  \mathcal{C}_p \xi + \mathcal{D}_p  \epsilon_\parallel \xi^{3/2} + \mathcal{E}_p  \xi^2 \right),
	\end{split}
\end{equation}
where we define the terms
\begin{align}
	\mathcal{A}_p &= \left(1 - \frac{U_\parallel^2}{v_T^2}\right) \left(A_n - \frac{3}{2} A_T - z^2 F_p \right) - \frac{U_\parallel}{v_T} \left(2 A_U - \frac{U_\parallel}{v_T} A_T\right),\\
	\mathcal{B}_p &= \left(A_U \left(2 - 6 \frac{U_\parallel^2}{v_T^2} \right) + U_\parallel \left(2 A_n - 5 A_T - 2 z^2 F_p \right)\right) \lambda_1,\\
	\mathcal{C}_p &= \left(1 - \frac{U_\parallel^2}{v_T^2}\right) A_T + 4 A_U \frac{U_\parallel}{v_T} \lambda_2 + \frac{U_\parallel^2}{v_T^2} \lambda_2 \left(2 A_n - 7 A_T - 2 z^2 F_p \right),\\
	\mathcal{D}_p & = 2 A_T \frac{U_\parallel}{v_T} \lambda_1 + 4 A_U \frac{U_\parallel^2}{v_T^2} \lambda_3, \\
	\mathcal{E}_p & = 2 A_T \frac{U_\parallel^2}{v_T^2} \lambda_2, 
\end{align}
and where
\begin{equation}
	z^2 = \frac{\overbar{\omega}}{n \omega_{d0} F_p\left(k_+ d\right)}.
\end{equation}
Thus, the integral simplifies to
\begin{equation}
	I_{p,E} = \sum_{\epsilon_\parallel} 2 f_p \int_0^\infty d\xi \left(2 \pi\right)^2 \frac{n_0 e^2}{T} \frac{R_0 r_0}{\sqrt{\pi}} \sqrt{\xi} \frac{ \mathcal{A}_p + \mathcal{B}_p  \epsilon_\parallel \xi^{1/2}  +  \mathcal{C}_p \xi + \mathcal{D}_p  \epsilon_\parallel \xi^{3/2} + \mathcal{E}_p  \xi^2 }{F_p\left(k_+ d\right) \left( \xi + \epsilon_\parallel \sqrt{\xi} \frac{x_+}{d} \frac{v_T}{q R_0 F_p\left(k_+ d\right)} - z^2\right)}.
\end{equation}
We then perform a change in variables to $u = \sqrt{\xi}$ and note that
\begin{equation}
	\sum_{\epsilon_\parallel} \int_0^{\infty} d \xi \sqrt{\xi} g\left(\epsilon_\parallel \sqrt{\xi}\right) = \sum_{\epsilon_\parallel} \int_0^{\infty}du  2 u^2 g\left(\epsilon_\parallel u\right) = \int_{-\infty}^{\infty} du 2 u^2 g\left(u\right).
\end{equation}
The integral then becomes
\begin{equation}
	\begin{split}
	I_{p,E} = 4 f_p \int_{-\infty}^\infty d u \left(2 \pi\right)^2 \frac{n_0 e^2}{T} \frac{R_0 r_0}{\sqrt{\pi}} u^2 \frac{ \mathcal{A}_p + \mathcal{B}_p  u  +  \mathcal{C}_p u^2 + \mathcal{D}_p  u^3 + \mathcal{E}_p  u^4}{F_p\left(k_+ d\right) \left( u^2 + u \frac{x_+}{d} \frac{v_T}{q R_0 F_p\left(k_+ d\right)} - z^2\right)}. 
	\end{split}
\end{equation}
To simplify this integral further, we rewrite the denominator as
\begin{equation}
	u^2 + u \frac{x_+}{d} \frac{v_T}{q R_0 F_p\left(k_+ d\right)} - z^2 = \left(u - z_+\right) \left(u - z_-\right),
\end{equation}
where
\begin{equation}
	z_{\pm} = -\frac{1}{2} \frac{x_+}{d} \frac{v_T}{q R_0 F_p\left(k_+ d\right)} \pm \sqrt{\left(\frac{1}{2} \frac{x_+}{d} \frac{v_T}{q R_0 F_p\left(k_+ d\right)}\right)^2 + z^2}. 
\end{equation}
This allows us to simplify the integral using the plasma dispersion functions defined in Appendix ~\ref{Fried and Conte}, allowing us to obtain
\begin{equation}
	\begin{split}
	I_{p,E} = 4 f_p \left(2 \pi\right)^2 \frac{n_0 e^2}{T}  \frac{R_0 r_0}{F_p\left(k_+ d\right)} \left( \mathcal{A}_p G_2  + \mathcal{B}_p  G_3  +  \mathcal{C}_p G_4  + \mathcal{D}_p G_5 + \mathcal{E}_p  G_6 \right),
	\end{split}
\end{equation}
where the associated Fried and Conte integrals $G_n = G_n\left(z_+, z_-\right)$ are evaluated at $z_+$ and $z_-$. Thus, the passing functional simplifies to
\begin{equation}
	\begin{split}
	\mathcal{L}_\text{passing} = \int_{-\infty}^{\infty} \int_{-\infty}^{\infty} &\frac{d \rho_\ast dk_\ast }{\sqrt{\pi}} \left(2 \pi\right)^3 \frac{e^2 n_0}{T} r_0 R_0 f_p  \left<\mathcal{I}_p\right>_{\xi, \lambda} \Gamma_0 \left(k_\perp \rho_{\text{th}}\right) \\ & \times \sqrt{\frac{\Imag\left(w^2\right)^2}{\Real\left(w^2\right)} + \Real\left(w^2\right)} e^{-\rho_\ast^2} e^{-k_\ast^2} \exp\left(\frac{\Imag\left(x_0\right)^2}{\Real\left(w^2\right)}\right), 
	\end{split}
\end{equation}
where
\begin{equation}
	\left<\mathcal{I}_p\right>_{\xi, \lambda} = \frac{2}{\pi F_p \left(k_+\right)}\left(\mathcal{A}_p G_2 + \mathcal{B}_p  G_3  +  \mathcal{C}_p G_4 + \mathcal{D}_p  G_5 + \mathcal{E}_p  G_6 \right). 
\end{equation}
We have now reduced all parts of the dispersion relation to a numerically tractable form. The adiabatic piece can be calculated analytically, whereas the trapped and passing functionals require $1$- and $2$-dimensional integrals, respectively. With the dispersion relation in hand, we can proceed to applying quasilinear theory.

\section{Quasilinear Approximation} \label{Quasilinear}
The core principle of quasilinear theory is to consider the slow time variation of the total distribution function $f$ and the resultant fluxes that attempt to drive the distribution function back to equilibrium. The validity of the quasilinear approximation depends on the decorrelation time of the potential being shorter than the eddy turn-over time. The ratio of these two quantities is known as the Kubo number \citep{kubo1963, krommes2002}. The single particle analogue to this is that the individual particle must not be trapped in the field; this allows the dynamics to be characterized as a random walk process, leading to a justification for the quasilinear approach. These characteristic times have been calculated and compared for both ETG and ITG-TEM turbulence \citep{lin2008, casati2009, citrin2012}. For these general cases, the Kubo number is less than unity and well developed turbulence for tokamak plasma parameters manifests random walk processes. Moreover, it has been found that quasilinear models are successful in reproducing experimental results such as temperature profiles within $15\%$ rms error \citep{kinsey2008}.

To proceed, we first recall the Vlasov equation for a given species (again omitting the species label):
\begin{equation}
\frac{\partial f}{\partial t} + \dot{\bm{\alpha}} \cdot \frac{\partial f}{\partial \bm{\alpha}} + \dot{\vec{J}} \cdot \frac{\partial f}{\partial \vec{J}} = 0.
\end{equation}
When we obtained the dispersion relation, we considered the linear response and neglected terms that are quadratic in the fluctuations. Moreover, we also assumed $f_0$ was time independent. To proceed with the quasilinear approximation, we now suppose that $f_0$ varies slowly in time on a time scale longer than that of the linear modes. We may then perform a time average over the Vlasov equation such that $\left<f\right>_t = f_0$ and the linear response averages to zero. We define the time average as
\begin{equation}
	\left<g\left(t\right)\right>_t = \frac{1}{T} \int_{-T/2}^{T/2} g\left(t + t'\right) dt',
\end{equation}
where $T$ is the time scale associated with the linear modes. The time averaged Vlasov equation then reads
\begin{equation}
	\left<\frac{\partial f}{\partial t} + \dot{\bm{\alpha}} \cdot \frac{\partial f}{\partial \bm{\alpha}} + \dot{\vec{J}} \cdot \frac{\partial f}{\partial \vec{J}}\right>_t \approx \frac{\partial f_0}{\partial t} + \left<\left\{\Real(\delta f), \Real(e \phi) \right\} \right>_t = 0. 
\end{equation}
Here, we take the real part of $\delta f$ or $\phi$ to obtain the physical quantity in accordance with our convention. To proceed, we rewrite the Poisson bracket as
\begin{equation}
	\left\{\Real(\delta f), \Real(e \phi) \right\}  = \frac{\partial}{\partial \bm{\alpha}} \cdot \left(\Real(\delta f) \frac{\partial \Real(e \phi)}{\partial \vec{J}}\right) - \frac{\partial}{\partial \vec{J}} \cdot \left(\Real(\delta f) \frac{\partial \Real(e \phi)}{\partial \bm{\alpha}}\right).
\end{equation}
The time average can be simplified by noting that for any two general vectors $\vec{A}$ and $\vec{B}$ we have
\begin{equation}
	\left<\Real\left(\vec{A} e^{- i \omega t}\right) \cdot \Real\left(\vec{B} e^{- i \omega t}\right)\right>_t = \frac{1}{2} \Real\left(\vec{A} \cdot \vec{B}^\ast\right). 
\end{equation}
Due to the Fourier structure of $\delta f$ and $\phi$, we also note that 
\begin{equation}
	\frac{\partial}{\partial \bm{\alpha}} \left< \Real\left(\delta f\right) \Real\left(e \phi\right) \right>_t = \vec{0}. 
\end{equation}
Essentially, the $\bm{\alpha}$ dependence disappears after performing the time average. Moreover, taking the real part of $\delta f$ and $\phi$ commutes with taking derivatives of real variables. We therefore obtain
\begin{equation}
	\frac{\partial f_0}{\partial t} + \frac{\partial}{\partial \vec{J}} \cdot \vec{\Gamma}_Q = 0, 
\end{equation}
where we define the quasilinear flux $\vec{\Gamma}_Q$ as
\begin{equation}
	\begin{split}
	\vec{\Gamma}_Q = \frac{1}{2} \Real\left(\sum_{\vec{n}} i \vec{n} f_{\vec{n}} e \phi_{\vec{n}}^\ast \right) = -\frac{1}{2} \Imag \left(\sum_{\vec{n}} \vec{n} \frac{e^2 \left|\phi_{\vec{n}}\right|^2}{T} f_0 \left(1 - \frac{\omega - \vec{n} \cdot \vec{\omega}_\ast - \vec{n} \cdot \vec{\omega}_E}{\omega - \vec{n} \cdot \vec{\Omega}}\right) \right). 
	\end{split}
\end{equation}
Here, $f_{\vec{n}}$ and $\phi_{\vec{n}}$ are related via the dispersion relation in the linearized problem. Thus, the quasilinear flux is computed by substituting in the solution of the dispersion relation including the found eigenvalues $\omega$, again only considering unstable modes. Modes that lack unstable solutions do not contribute to the quasilinear flux. 

We are now in a position to calculate the flux surface averaged particle, toroidal angular momentum, and energy fluxes by averaging the Vlasov equation over velocity and space. This is analogous to calculating the fluid equations by taking moments of the Vlasov equation. The radial fluxes can be calculated via a change in variables from $J_3$ to $r$. We find that
\begin{align}
	\frac{\partial \left<n\right>}{\partial t} + \frac{d \Gamma}{dr} &= 0,\\
	\frac{\partial \left<m n R U_\parallel\right>}{\partial t} + \frac{d \Pi}{dr} &= 0,\\	
	\frac{3}{2} \frac{\partial \left<p \right>}{\partial t} + \frac{d Q}{dr} & = 0,
\end{align}
where $\Gamma$, $\Pi$, and $Q_E$ are the particle, toroidal momentum, and energy fluxes defined as
\begin{align}
	\Gamma &= \frac{1}{4 \pi^2 d} \int d^3v d^3r   \frac{1}{2}  \Imag\left(\sum_{\vec{n}} \frac{n q}{r B} f_{\vec{n}} \phi_{\vec{n}}^\ast \right),\\
    \Pi &= \frac{1}{4 \pi^2 d} \int d^3v d^3r \frac{m R v_\parallel}{2}  \Imag\left(\sum_{\vec{n}} \frac{n q}{r B} f_{\vec{n}} \phi_{\vec{n}}^\ast \right),\\
	Q_E &= \frac{1}{4 \pi^2 d} \int d^3v d^3r \frac{m\left(v^2 - U_\parallel^2\right)}{4}  \Imag\left(\sum_{\vec{n}} \frac{n q}{r B} f_{\vec{n}} \phi_{\vec{n}}^\ast \right). 
\end{align}
Here, we can see that the integrations to calculate the particle, toroidal momentum, and energy fluxes are of the same form to solve the dispersion relation. The particle flux calculation is identical. Meanwhile, we must take into account an extra factor of $v_\parallel$ and $v^2$ for the angular momentum flux and energy flux integrations, respectively. These changes can be easily accommodated for without affecting the fundamental approach. For instance, the inclusion of $v^2$ simply changes the associated Fried and Conte integral. The physical significance of these fluxes can be further solidified by examining the perturbed $E$-cross-$B$ velocity. We find that
\begin{equation}
	\delta \vec{v}_{E\times B} \cdot \hat{r} = \hat{r} \cdot -\frac{\nabla \delta \phi \times \vec{B}}{B^2} \approx \sum_{\vec{n}} \frac{i k_\theta \phi_{ \vec{n}}}{B} = \sum_{\vec{n}} \frac{i n q}{r B} \phi_{\vec{n}},
\end{equation}
where we have again used the convention that $k_\theta \to \left(i/r\right) \partial_\theta $. We then find that
\begin{equation}
	\left< \Real\left(\delta \vec{v}_{E\times B} \cdot \hat{r}\right) \Real\left(\delta f\right) \right>_t = \frac{1}{2} \Real\left( \sum_{\vec{n}} f_{\vec{n}} \left(\frac{i n q}{r B} \phi_{\vec{n}} \right)^\ast \right) = \frac{1}{2} \Imag\left( \sum_{\vec{n}} f_{\vec{n}} \frac{n q}{rB} \phi_{\vec{n}}^\ast \right).
\end{equation}
This lets us write the fluxes as
\begin{align}
	\Gamma &= \left< \delta \left(n\right) \delta \vec{v}_{E\times B} \cdot \hat{r} \right>_{t, r},\\
	\Pi &= \left< \delta \left(m n R U_\parallel\right) \delta \vec{v}_{E\times B} \cdot \hat{r} \right>_{t, r},\\
	Q_E &= \left< \delta \left(P\right)  \delta \vec{v}_{E\times B} \cdot \hat{r} \right>_{t, r}.
\end{align}
Therefore, the particle, angular momentum, and energy fluxes are simply related to moments of the perturbed distribution function integrated against the perturbed $E$-cross-$B$ velocity, where $\left<\dots\right>_{t,r}$ denotes a time and spatial average. We also define
\begin{align}
	\delta n  &= \int d^3v \delta f,\\
	\delta \left(n m R U_\parallel \right) &= \int d^3v m v_\parallel \delta f,\\
	\delta P & = \int d^3v \frac{1}{2} m \left(v^2 - U_\parallel^2\right)\delta f,
\end{align}
where we calculate the particle, angular momentum, and energy fluxes for every species. We note that the toroidal angular momentum flux is only non-zero in the presence of rotations. The energy flux calculation can be approximated by noting in the small Mach number limit that
\begin{align}
	\delta P = \delta \left(n T\right) = - \frac{1}{2} m U_\parallel^2 \delta n + \int d^3v \frac{1}{2} m v^2 \approx \int d^3v \frac{1}{2} m v^2. 
\end{align}
We also note that often we are concerned with the heat flux $Q$ relative to the convective energy flux $\frac{3}{2} T \Gamma$ \citep{horton1984}. The heat flux is simply
\begin{equation}
	Q = Q_E - \frac{3T}{2} \Gamma. 
\end{equation}

It is important to note that while we may obtain quasilinear flux ratios from the above procedure, we cannot with linear physics alone obtain the physical fluxes. Throughout the derivation, we have kept the amplitude of the fluctuating potential $\delta \phi$ arbitrary. The amplitude $\phi_0$ can only be obtained through the use of nonlinear physics by saturating the amplitude. Thus, the complete calculation of these fluxes must be obtained via a saturation rule obtained from a nonlinear computational code, in this case from the Gyrokinetic Electromagnetic Numerical Experiment (GENE) \citep{jenko2000}. This saturation rule is the topic of the next section.

\section{Saturation Rule} \label{Saturation}
To formulate a saturation rule, we introduce the well known mixing length estimate with an effective diffusivity $D$:
\begin{equation}
	D = \left. \frac{\gamma_n}{\left<k_\perp^2\right>} \right|_{\text{max}},
\end{equation}
where we compute the value of $\gamma_n$ such that the quantity $\gamma_n / \left<k_\perp^2\right>$ is at its maximum over the linear spectrum for a given mode. Meanwhile, we average $k_\perp^2$ over the electrostatic mode. We enforce this mixing length estimate for our various flux calculations by approximating the underlying process as a random walk \citep{bourdelle2007}. For instance, we mandate that the particle flux for a given species must be
\begin{equation}
	\Gamma_s = \sum_{n} C_{\text{NL}} \frac{S_n}{R_0 n_{0s}} 	\left. \frac{\gamma_n}{\left<k_\perp^2\right>} \right|_{\text{max}} \frac{k_\theta}{k_{\theta, \text{max}}} L_{s, n, 0},
\end{equation}
where $C_{\text{NL}}$ is a dimensionless constant from nonlinear physics, the form factor $S_n$ is a mode-dependent form factor, $k_{\theta, \text{max}}$ corresponds to the mode that maximizes $\gamma_n / k_\perp^2$, and $L_{s,n,0}$ is the dimensionless integral that actually computes the flux terms. The above expression is only valid when there is only one mode present in the linear spectrum. We can generalize the expression to account for the existence of multiple types of linear modes by introducing another form factor $S_{n'}$ into the expression and summing over both $n$ and $n'$, while we compute the maximum $\gamma_n / k_\perp^2$ for a given $n'$. 

We model $C_{\text{NL}}$ with the use of nonlinear gyrokinetic simulations. We distinguish between ITG scales, which we define as $k_\theta \rho_s < 2$, and ETG scales, which we define as $k_\theta \rho_s > 2$. Here, $\rho_s$ is the gyroradius of the main ion species such that $\rho_s = \sqrt{T_s / m_s} / \Omega_{1,s}$ (note that lack of $\sqrt{2}$). The ITG scales are tuned to the GA-Standard nonlinear ion heat flux computed by GENE, whereas the ETG scales are tuned to a single-scale nonlinear GENE simulation based on JET parameters \citep{citrin2017}. These parameters are current as of QuaLiKiz version 2.8.1 and are subject to future change depending on updates to the nonlinear physics. The result is
\begin{equation}
	C_{\text{NL}} =  \left\{ \begin{aligned}
		& 271 / s_{\text{fac}}  \qquad \qquad \quad && \text{ if } k_\theta \rho_s < 2 \text{ (ITG)},\\
		& 122 f_{\text{multi-scale}}/ s_{\text{fac}} && \text{ if }  k_\theta \rho_s > 2 \text{ (ETG)}.
	\end{aligned} \right.
\end{equation}
Here, we have also introduced an ad hoc factor $s_{\text{fac}}$ for the case of low magnetic shear \citep{citrin2012},
\begin{equation}
		s_{\text{fac}} = \left\{ \begin{aligned}
		& 2.5 \left(1 - \left|s\right|\right) \qquad && \text{ if } \left|s\right| < 0.6,\\
		& 1  && \text{ if }  \left|s\right| > 0.6,
	\end{aligned} \right.
\end{equation}
as well as a multi-scale rule determined from the maximum of the respective spectra,
\begin{equation}
	f_{\text{multi-scale}} = \frac{1}{1 + \exp\left(-\frac{1}{5} \left(	\frac{\gamma_{\text{ETG, max}}}{\gamma_{\text{ITG, max}}} - \sqrt{\frac{m_i}{m_e}}\right)\right)},
\end{equation}
where $m_e$ and $m_i$ are the masses of the electron and main ion respectively. Here, the sigmoid guarantees a smooth transition from a strongly driven ion-scale mode regime and a strongly driven electron-scale mode regime, since it has been observed that ETG turbulence is suppressed when the ion-scale instability dominates. 

Lastly, we provide an explicit expression for $k_\perp^2$. In the ITG regime, we need to take into account contributions to $k_r^2$ that arise from the magnetic shear, the mode structure of the electrostatic perturbation, and nonlinear effects. Meanwhile, in the ETG regime we assume full isotropization of the mode such that $k_r^2 = k_\theta^2$. The result is 
\begin{equation}
	\left<k_\perp^2\right> = \left\{ \begin{aligned}
		& k_\theta^2 + \left(k_{r-\text{NL}} + k_{r-\text{shear}}\right)^2  \qquad  && \text{ if } k_\theta \rho_s < 2 \text{ (ITG)},\\
		& 2 k_\theta^2 && \text{ if }  k_\theta \rho_s > 2 \text{ (ETG)}.
	\end{aligned} \right.
\end{equation}
The shear contribution can be calculated analytically as
\begin{equation}
	k_{r-\text{shear}} = k_\theta \left|s\right| \sqrt{\left<\theta^2\right>} = \frac{k_\theta s d}{\sqrt{2} \Real\left(w^2\right)} \sqrt{\Real\left(w^2\right) + 2 \Imag\left(x_0\right)^2},
\end{equation}
where we use
\begin{equation}
	\left<\theta^2\right> = \frac{\int_{-\infty}^{\infty} \theta^2 \left|\oldhat{\phi}\left(\theta\right)\right|^2 d\theta }{\int_{-\infty}^{\infty} \left|\oldhat{\phi}\left(\theta\right)\right|^2 d\theta} = \frac{d^2}{2 \Real\left(w^2\right)} + \left(d \frac{\Imag\left(x_0\right)}{\Real\left(w^2\right)}\right)^2. 
\end{equation}
Meanwhile, the nonlinear contribution has been tuned \citep{citrin2012} such that
\begin{equation}
	k_{r-\text{NL}} \rho_s = 0.4 e^{-2 \left|s\right|} q^{-0.5} + 1.5 \max\left\{k_\theta \rho_s - 0.2, 0\right\}. 
\end{equation}

Having now fully derived analytic expressions for the dispersion relation and quasilinear fluxes, we now discuss the numerical implementation of QuaLiKiz. 

\section{Numerical Implementation} \label{Numerical}
Recall that the dispersion relation is written as
\begin{equation}
	\sum_s \mathcal{L}_{0, s} - \mathcal{L}_{\text{passing}, s} - \mathcal{L}_{\text{trapped}, s} = 0.
\end{equation}
The trapped and passing functionals discussed in Sections~\ref{Trapped} ~and~\ref{Passing} are both functions of the complex frequency $\omega$. Solving the dispersion relation is therefore a matter of finding the zeros of the complex analytic function $D\left(\omega\right)$, where 
\begin{equation}
	D\left(\omega\right) = \sum_s \mathcal{L}_{0, s} - \mathcal{L}_{\text{passing}, s} - \mathcal{L}_{\text{trapped}, s}.
\end{equation}
To solve this, we use the Davies method, a numerical technique developed by \citet{davies1986} to find the zeros of an analytic function within the complex plane. The strategy takes advantage of the argument principle in complex analysis, which states that given a meromorphic function $f\left(z\right)$ that
\begin{equation}
	\frac{1}{2 \pi i} \oint_C \frac{f'\left(z\right)}{f\left(z\right)} dz = N - P,
\end{equation}
where $N$ and $P$ are respectively the number of zeros and poles of $f\left(z\right)$ contained within the simple counter-clockwise contour $C$. Here, zero multiplicity and pole order are taken into account. For our purposes, we assume that $f\left(z\right)$ has no poles, leading to
\begin{equation}
	\frac{1}{2 \pi i} \oint_C \frac{f'\left(z\right)}{f\left(z\right)} dz = N. 
\end{equation}
The key of the method is to recognize from Cauchy's residue theorem that, for integer $n$ such that $1 \le n \le N$, we can calculate the integral $S_n$ such that
\begin{equation}
	S_n = \frac{1}{2 \pi i} \oint_C z^n \frac{f'\left(z\right)}{f\left(z\right)} dz = \sum_{j=1}^{N} z_{0j}^n,
\end{equation}
where $z_{0j}$ is the $j$th root of $f\left(z\right)$ (counting repeated roots as separate). We then construct the polynomial
\begin{equation}
	P_N\left(z\right) = \prod_{j=1}^{N} \left(z - z_{0j}\right) = \sum_{j=0}^N A_j z^{N-j},
\end{equation}
where the coefficients $A_j$ can be computed from the relations
\begin{align}
	A_0 & = 1,\\
	S_1 + A_1 &= 0,\\
	S_2 + A_1 S_1 + 2 A_2 &= 0,\\
	S_{n} + A_1 S_{n-1} + A_2 S_{n-2} + \dots + n A_n  &= 0, \qquad n = 1,2,\dots,N,
\end{align}
Excluding the trivial $A_0$ term, this is a linear system of $N$ equations. After solving this system, we can then construct the polynomial $P_N$ which has zeros that are precisely the solutions of the dispersion relation. We then extract a zero from the polynomial $P_N$ using a Newton solver and then define a new set of coefficients such that
\begin{equation}
	S^{\left(1\right)}_n = S_n - z_{01}^n, \qquad n = 0, 1, \dots, N-1 \\
\end{equation}
where $z_{01}$ is the first zero found. With this new set of coefficients, we may then construct a new polynomial $P_{N-1}\left(z\right)$ and extract another zero. This process is repeated until all zeros are found. If the contour of integration is a unit circle, then a clever integration by parts results in
\begin{equation}
	S_n = -\frac{n}{2 \pi i} \int_{0}^{2\pi} d\theta e^{i n \theta} \ln \left(e^{- i N \theta} f\left(e^{i \theta}\right)\right), \qquad n > 0.
\end{equation}
The inclusion of $e^{- i N \theta}$ inside the logarithm is to handle the branch cut of the logarithm, and can be obtained by using $z^{-N} f\left(z\right)$ instead of $f\left(z\right)$ in the preceding formulas; such a substitution does not affect the value of $S_n$ for $n > 0$. For the $n= 0$ case, we simply compute the total change of the argument of $f\left(e^{i \theta}\right)$ for $0 \le \theta \le 2 \pi$ while keeping track of any jumps in the argument that would indicate a full winding. Thus, $S_n$ can be computed via standard quadrature methods for $1$-dimensional integration. 

To apply this to the dispersion relation, we make use of a bijective mapping $\omega = \omega\left(z\right)$ (to be determined momentarily). This will allow us to retain the simplifications that come from integrating around a unit circle. The first step is to define $f\left(z\right)$ such that
\begin{equation}
	f\left(z\right) = D\left(\omega\left(z\right)\right). 
\end{equation}
Then, we compute $S_n$ via numerical quadrature, leading to roots $z_{0n}$ such that 
\begin{equation}
	f\left(z_{0n}\right) = D\left(\omega\left(z_{0n}\right)\right) = 0
\end{equation}
Because the mapping is bijective, we may then simply apply the mapping onto the roots $z_{n0}$ to obtain
\begin{equation}
	\omega_{0n} = \omega\left(z_{0n}\right),
\end{equation}
where $\omega_{0n}$ are all the roots within the contour $C$ in the complex $\omega$-plane such that
\begin{equation}
	D\left(\omega_{0n}\right) = 0.
\end{equation}
The only task remaining is to define a suitable bijective mapping $\omega\left(z\right)$. Because QuaLiKiz only considers unstable modes, we demand that $\Imag\left(\omega\right) > 0$ along the entirety of the contour in the $z$-plane. We first define the bijective mapping $\left(u, v\right) \to \left(x, y\right)$ as
\begin{align}
	x\left(u, v\right) &= \frac{\text{sgn} \left(uv\right) }{v \sqrt{2}} \sqrt{u^2 + v^2 - \sqrt{\left(u^2 + v^2\right) \left(u^2 + v^2 - 4 u^2 v^2 \right)} },\\
	y\left(u, v\right) &= \frac{\text{sgn} \left(uv\right) }{u \sqrt{2}} \sqrt{u^2 + v^2 - \sqrt{\left(u^2 + v^2\right) \left(u^2 + v^2 - 4 u^2 v^2 \right)} }.
\end{align}
The inverse mapping is given by
\begin{align}
	u\left(x, y\right) &= \frac{x \sqrt{x^2 + y^2 - x^2 y^2}}{\sqrt{x^2 + y^2}}, \\
	v\left(x, y\right) &= \frac{y \sqrt{x^2 + y^2 - x^2 y^2}}{\sqrt{x^2 + y^2}}. 
\end{align}
Since this mapping does not satisfy the Cauchy-Riemann equations, it is merely bijective, not conformal. This is known as a squircle mapping since it appears to be a square with rounded edges, and this specific kind was first formulated in \citet{guasti1992}. Denoting $\omega = x' + i y'$ and $z = u + i v$, we modify this mapping such that
\begin{align}
	x' &= R_{x} + \frac{r_x}{a} x\left(a u, a v\right),\\
	y' &= R_{y} + \frac{1}{a} \left(R_{y} - \epsilon_y \right) y \left(a u, a v\right),
\end{align}
With $C$ being the unit circle in the complex $z$-plane, let $C'$ be the mapped curve in the complex $\omega$-plane. Here, $\left(R_x, R_y\right)$ determines the approximate center $C'$, $r_x$ and $a$ are scaling factors chosen to manipulate $C'$ into a rectangular shape, and $\epsilon_y$ is chosen to guarantee that $C'$ lies above the real axis. While the mapping is not conformal, it is sufficient for our method, since not only is it bijective but points interior to $C$ are mapped to the interior of $C'$. Thus, if we make the interior area of $C'$ sufficiently large and place it slightly above the real axis in the complex $\omega$-plane, then we will determine all eigenmodes of interest to us. After the solution frequencies are found, they are then refined using a standard Newton root-finding method. 

While the contour integral and the Newton root-finding are done when QuaLiKiz is used on its own, when coupled to an integrated modeling suite a slight modification is made to algorithm. We assume that the quasilinear transport changes slowly compared to the timescale of evolution of the plasma equilibrium. A typical transport solver iterates on a time step that is on the order of $\lesssim 10^{-2} \tau_E$, where $\tau_E$ is the energy confinement time. To speed up the code, QuaLiKiz will often only use the previous solution as an initial guess for the Newton solver rather than perform the full contour integral. Since codes like QuaLiKiz are often the bottleneck for the whole integrated modeling suite, such a speedup is necessary to make the simulation tractably feasible. In practice, QuaLiKiz will only perform the full contour integral once every $\sim 10$ iterations.

Lastly, we discuss the numerical integration scheme currently in use by QuaLiKiz to calculate the trapped and passing functionals, which require $2$-dimensional integrations. Although QuaLiKiz used to rely on integration routines provided by the Numerical Algorithms Group (NAG), it now uses open source routines based the Genz and Malik algorithm, dubbed ``hcubature''. This algorithm was originally developed by \citet{genz1980}; the current implementation is based on the C++ implementation \citet{johnson2017}. The version of the algorithm in QuaLiKiz has been ported to Fortran and is slightly modified as a result. 

The goal of hcubature is to estimate 
\begin{equation}
	\vec{I} = \int_{a_1}^{b_1} \int_{a_2}^{b_2} \dots \int_{a_n}^{b_n} \vec{f}(\vec{x}) \text{d}^n x.
\end{equation}
Here, $\vec{I}$ is the estimate of the integral, while $a_i$ and $b_i$ are respectively the individual components of the lower and upper bounds of the integral $\vec{a}$ and $\vec{b}$, which are both constant vectors with dimension $n$. Meanwhile,  $\vec{f}$ is a vector function of arbitrary dimension, and $\vec{x}$ is the argument of the function $\vec{f}$ and is of dimension $n$. The vectors $\vec{I}$ and $\vec{f}$ are of the same dimension. Thus, hcubature approximately integrates a vector integrand over a hyperrectangle (or equivalently a scaled hypercube, hence the name ``cubature''). The routine terminates when the global estimate of the absolute or relative error of the integral reach the desired tolerance and also calculates an error vector $\bm{e}$ with the same dimensionality as the integrand. While calculating the error vector is straightforward, incorporating it into the convergence criterion is non-trivial. In general, to estimate the error, we make a higher order estimate $\vec{I}_0$ and a lower order, less accurate estimate $\vec{I}_1$ and set the $i$th component of $\bm{\epsilon}$ to be
\begin{equation}
	\epsilon_i = \left|(I_0)_i - (I_1)_i\right|.
\end{equation}

For simplicity, we first consider a scalar function that we integrate over a hypercube,
\begin{equation}
	I = \int_{-1}^1 \int_{-1}^1 \dots \int_{-1}^1 f(x_1, x_2, \dots, x_n) \text{d}^n x.
\end{equation}	
We estimate the integral using the following rule,
\begin{equation}
	\begin{split}
		I \approx I_0 =  & w_1 f\left(0, 0, \dots , 0\right) + w_2 \sum f\left(\lambda_2, 0, 0,\dots, 0\right) + w_3\sum f\left(\lambda_3, 0, 0,\dots,0\right) \\ &+ w_4 \sum f\left(\lambda_4, \lambda_4, 0, 0,\dots,0\right) + w_5 \sum f\left(\lambda_5, \lambda_5,\dots,\lambda_5\right).
	\end{split}
\end{equation}
Here, we sum over all possible permutations of coordinates while also allowing for sign changes. For example, if $f$ takes three arguments, then
\begin{equation}
	\begin{split}
		\sum f\left(\lambda_1, 0,0\right) = f\left(\lambda_1,0,0\right) + &f\left(0,\lambda_1,0\right) + f\left(0,0,\lambda_1\right) \\ &+  f\left(-\lambda_1,0, 0\right) + f\left(0,-\lambda_1,0\right) + f\left(0,0,-\lambda_1\right).
	\end{split}                                                                                                           
\end{equation}                                                                                                         
Genz and Malik constrain the parameters $w_i$ and $\lambda_i$ by requiring that the integration be exact for the functions
\begin{equation}
	\begin{split}
		&f_1\left(x_1, x_2,\dots,x_n\right) = 1,\\
		&f_2\left(x_1, x_2,\dots,x_n\right) = x_1^2,\\
		&f_3\left(x_1, x_2,\dots,x_n\right) = x_1^4,\\
		&f_4\left(x_1, x_2,\dots,x_n\right) = x_1^6,\\
		&f_5\left(x_1, x_2,\dots,x_n\right) = x_1^2 x_2^2,\\
		&f_6\left(x_1, x_2,\dots,x_n\right) = x_1^4 x_2^2,\\
		&f_7\left(x_1, x_2,\dots,x_n\right) = x_1^2 x_2^2 x_3^2.
	\end{split}
\end{equation}
In addition, they also fix the parameters $\lambda_3 = \lambda_4$ to be a specific number, and solve the resulting nonlinear system of equations. The result can be found in \citet{genz1980}. To estimate the error, we reuse $\lambda_i$ but calculate different weights $w'_i$ to make a lower-order estimate,	
\begin{equation}
	\begin{split}
	I \approx I_1 = w'_1 f\left(0, 0,\dots,0\right) + & w'_2 \sum f\left(\lambda_2, 0, 0,\dots,0\right) \\ &+ w'_3 \sum f\left(\lambda_3,0,0,\dots,0\right) + w'_4 \sum f\left(\lambda_4,\lambda_4,0,0,\dots,0\right).
	\end{split}
\end{equation}
We calculate the weights with the same method as previously discussed and require the integration be exact for the functions $f_1, f_2, f_3,$ and $f_5$. By keeping $\lambda_i$ the same, we can estimate the error without reusing function evaluations. The error is taken to be
\begin{equation}
	\epsilon = \left|I_0 - I_1\right|.
\end{equation}
The estimate procedure easily generalizes to that of a hyperrectangle by using linear transformations. The calculation of $I_0$, $I_1$, and $\vec{\epsilon}$ can also be extended to the case of vector integrands by integrating every component simultaneously. 

In the case that $n=1$, the above rule no longer applies. Instead, hcubature uses a 15-point Kronrod extension of a 7-point Gaussian quadrature rule. For $n$-point Gaussian quadrature, we estimate the integral
\begin{equation}
	I = \int_{-1}^1 f(x) \text{d}x \approx \sum_{i=1}^n w'_i f(x_i).
\end{equation}
To calculate the weights $w_i$ and the abscissa $x_i$, we require that the integration be exact for all polynomials up to degree $2 n - 1$. It can be shown using Lagrange interpolating polynomials and the theory of orthogonal polynomials that the abscissa $x_i$ correspond to the roots of the Legendre polynomial $P_n$ and that the weights are
\begin{equation}
	w'_i = \frac{2}{\left(1 - x_i^2\right)\frac{d P_n}{dx}\left(x_i\right)^2},
\end{equation}
where the Legendre polynomials are normalized such that $P_n(1) = 1$. 

One downside to this method is that the abscissa will in general be completely different for different order rules. Thus, naively comparing an $n$-point rule with an $n+1$-point rule to estimate the error is inefficient. Kronrod discovered that one for an $n$-point Gaussian quadrature rule, one could add $n+1$ abscissa to exactly integrate polynomials up to order $3n + 1$, reusing the previous abscissa and computing new weights $w_i$. These new nodes correspond to the zeros of Legendre-Stieltjes polynomials, and their derivation will not be covered here. Thus, the 15-point rule corresponds to
\begin{equation}
	I \approx I_0 = \sum_{i=1}^{15} w_i f(x_i),
\end{equation}
the 7-point rule to
\begin{equation}
	I \approx I_1 = \sum_{i=1}^7 w'_i f(x_i),
\end{equation}
and the estimated error
\begin{equation}
	\epsilon = \left|I_0 - I_1\right|.
\end{equation}
Extending this to more general limits of integration simply requires a linear transformation. 

Now that we have our integration schemes and error estimation rules for arbitrary $n$, we may proceed to describe the general algorithm. 
\begin{algorithm}
	\KwIn{$\vec{f}$, $\vec{a}$, $\vec{b}$, $\epsilon_a$, $\epsilon_r$, maxEval, norm}
	\KwOut{$\vec{I}$, $\vec{e}$, ifail}
	Initialize $\text{eval} = 0$\\
	Create a hyperrectangle from $\vec{a}$, $\vec{b}$\\
	Calculate $\vec{I}_0$, $\vec{I}_1$, $\bm{\epsilon}$ in the hyperrectangle\\
	Calculate $s$, the suggested dimension along which to further discretize, in the hyperrectangle\\
	Update eval to be the number of points evaluated so far\\
	$\vec{I} = \vec{I}_0, \vec{e} = \bm{\epsilon}$\\
	Push hyperrectangle into the binary heap with associated value $\max_{i} \left|\epsilon_i\right|$ and with stored values $\vec{I}_0, \bm{\epsilon}, s$\\
	\lIf{converged}{exit}
	\Else{
		\Repeat{converged or $\text{eval} \ge \text{maxEval}$}{
			Pop a hyperrectangle from the binary heap\\
			Update $\vec{I} = \vec{I} - \vec{I}_0$, $\vec{e} = \vec{e} - \bm{\epsilon}$ from the popped hyperrectangle\\
			Split the hyperrectangle in half along the suggested dimension $s$\\
			Calculate $\vec{I}_0$, $\vec{I}_1$, $\bm{\epsilon}$, and $s$ for each hyperrectangle\\
			Update eval to be the number of points evaluated so far\\
			Update $\vec{I} = \vec{I} + \sum \vec{I_0}, \vec{e} = \vec{e} + \sum \bm{\epsilon}$ from the two hyperrectangles\\
			Push each hyperrectangle into the binary heap with associated value $\max_{i} \left|\epsilon_i\right|$ and with stored values $\vec{I}_0, \bm{\epsilon}, s$\\
		}
	}
	
	\caption{hcubature}	
\end{algorithm}
Here, $\vec{f}$ is the vector integrand, $\vec{a}$ and $\vec{b}$ are respectively the lower and upper bounds of the integrand, $\epsilon_a$ and $\epsilon_r$ are respectively the requested absolute and relative error tolerances, maxEval is the maximum number of function evaluations to be allowed by the routine, and norm determines the convergence criterion (in conjunction with the requested error tolerances). The integer eval keeps track of the total number of function evaluations, the vectors $\vec{I}_0$ and $\vec{I}_1$ correspond to the integration estimates for a given hyperrectangle, $\bm{\epsilon}$ is the error estimate for the hyperrectangle, and $s$ is the suggested dimension of splitting. As for the output, $\vec{I}$ is the total integration estimate, $\vec{e}$ is the total error, and ifail is an integer denoting whether any errors occurred while carrying out the procedure or whether the eval reached maxEval before convergence. Convergence is determined using the global error vector $\vec{e}$. 

The algorithm splits the initial hyperrectangle into pieces and stores them in a binary heap. The heap is sorted according to the largest component of the local error vector, where the root of the heap corresponds to the region with the largest error. Until the integral converges, we pop a hyperrectangle from the root of the heap, split it into two regions, evaluate both regions accordingly, update the global integration and error estimates, and push both regions into the heap. This guarantees that the split region contributes the greatest to the global error. To determine which direction to split the hyperrectangle along, we calculate a fourth divided difference using the same evaluation points,
\begin{equation}
	\begin{split}
		D_i = \sum_j & \bigl|f_j(0,0,\dots,0,-\lambda_2,0,0,\dots,0) \\& - 2 f_j(0,0,\dots,0)  + f_j(0,0,\dots,0,\lambda_2,0,0\dots,0) \bigr. \\ & - \bigl.\frac{\lambda_1^2}{\lambda_2^2} \left[f_j(0,0,\dots,0,-\lambda_1,0,0,\dots,0) - 2 f_j(0,0,\dots,0) \right. \\ & \left. \qquad  + f_j(0,0,\dots,0,\lambda_1,0,0\dots,0)\right]\bigr|.
	\end{split}
\end{equation}
Here, $i$ corresponds to the dimension at which we evaluate the functions at. For example, if $i=2$, then
\begin{equation}
	\begin{split}
		D_2 = \sum_j & \bigl|f_j(0,-\lambda_2,0,0,\dots,0) - 2 f_j(0,0,\dots,0) + f_j(0,\lambda_2,0,0\dots,0) \bigr. \\ & - \bigl.\frac{\lambda_1^2}{\lambda_2^2} \left[f_j(0,-\lambda_1,0,0,\dots,0) - 2 f_j(0,0,\dots,0) + f_j(0,\lambda_1,0,0\dots,0)\right]\bigr|.
	\end{split}
\end{equation}
Note that here we take the difference along each component of $\vec{f}$ and sum the absolute value of each difference. We determine $s$, the dimension along which we split the hyperrectangle, by calculating the maximum component of $\vec{D}$. The coordinate corresponding to the maximum of $\vec{D}$ is the one in which we split the hyperrectangle in half. For the $1$-dimensional case using the Gauss-Kronrod rule, no such calculation is required. We continually split the whole hyperrectangle into smaller and smaller pieces until convergence is achieved.

\section{Conclusions and Outlook} \label{Conclusions}
In this work, we derived the linear dispersion relation of quasilinear gyrokinetic transport code QuaLiKiz from first principles. With the aid of nonlinear simulations, we also extended the linear physics to a quasilinear regime to calculate particle, toroidal angular momentum, and heat fluxes. The formulation of QuaLiKiz relies upon multiple theoretical principles in fusion plasma physics. First, we examined single particle motion in a circular magnetic geometry and identified the adiabatic invariants of motion within a guiding center framework. This allowed us to characterize electrostatic perturbations to the system with the aid of action-angle variables. We used this formulation to analyze the linearized Vlasov equation and Poisson's equation. We then simplified the resulting dispersion relation using the ballooning representation, an eigenfunction ansatz, and various approximations. The solution of this dispersion relation is computed using the Davies method and numerical cubature methods. Finally, upon finding the eigenmodes of the system, we use the solutions to compute the quasilinear fluxes with the aid of a saturation rule informed by nonlinear simulations. 

This derivation serves not only to help explain the inner workings of the model, but also to guide potential improvement in QuaLiKiz. With the formulation finally laid out, it is now clear where each individual approximation enters the derivation. This will ease future QuaLiKiz development that aims to extend the underlying physics or relax the various approximations. Examples of such work includes introducing electromagnetic perturbations, incorporating a more general magnetic geometry, and a more accurate pitch angle integration for passing particles. Improvements made to QuaLiKiz will allow for more accurate integrated modeling as well as further optimization of the code. 

An additional goal of this work is to provide an extensive, line-by-line derivation for the sake of demonstrating how such a model can be formulated in principle. Explicitly drawing upon multiple theoretical principles, such as the action-angle variable formalism, helps to illustrate the utility of these principles and their physical motivation. It is also useful to lay out the various mathematical and numerical techniques necessary in a model such as this, since many such tricks, methods, or approximations are often crucial in making a problem computationally tractable. We hope that this work will function not just as a tutorial for understanding and improving QuaLiKiz, but also further development in quasilinear fusion codes in general.

\appendix

\section{Fried and Conte Integrals} \label{Fried and Conte}
The Fried and Conte integral, also known as the plasma dispersion function, is utilized frequently in kinetic plasma physics. It is defined as
\begin{equation}
	Z(x) = 
	\begin{cases}
		\frac{1}{\sqrt{\pi}} \int_{-\infty}^{\infty} dv \frac{e^{-v^2}}{v-x}, & \text{if}\ \Imag(x) > 0,\\
		\mathcal{P} \frac{1}{\sqrt{\pi}} \int_{-\infty}^{\infty} dv \frac{e^{-v^2}}{v-x} + \sqrt{\pi} i e^{-x^2}, & \text{if}\ \Imag(x) = 0,\\
		\frac{1}{\sqrt{\pi}} \int_{-\infty}^{\infty} dv \frac{e^{-v^2}}{v-x} + 2\sqrt{\pi} i e^{-x^2}, & \text{if}\ \Imag(x) < 0,\\
	\end{cases}
\end{equation}
where the case $\Imag(x) \le 0$  is calculated by analytically continuing the integral defined for $x > 0$. When solving the Vlasov equation as an initial value problem in time, a Laplace transform is implied when obtaining this integral. To apply the Laplace transform correctly for the case of stable modes, we must analytically continue the function. Luckily, since we only consider unstable modes, we are free to restrict ourselves instead to the related function
\begin{equation}
	Z_0(x) = \frac{1}{\sqrt{\pi}} \int_{-\infty}^\infty dv \frac{e^{-v^2}}{v-x}. 
\end{equation}
If $\Imag(x) = 0$, we take the Cauchy principle value of $Z_0$. 

In carrying out the calculation, we define a generalization of the plasma dispersion function defined as
\begin{equation}
	Z_m(x) = \frac{1}{\sqrt{\pi}} \int_{-\infty}^{\infty} dv \frac{v^m e^{-v^2}}{x-v},
\end{equation} 
where $m$ is a nonnegative integer. It can be shown that these associated Fried and Conte integrals can be written in terms of $Z_0(x)$:
\begin{equation}
	Z_m(x) = \begin{cases}
		x^m Z_0(x) + \frac{1}{\sqrt{\pi}} \sum_{k=0}^{\frac{m-1}{2}} x^{2k} \Gamma(\frac{m}{2}-k), & \text{if m odd},\\
		x^m Z_0(x) + \frac{1}{\sqrt{\pi}} \sum_{k=0}^{\frac{m}{2}-1} x^{2k+1} \Gamma(\frac{m-1}{2}-k), & \text{if m even},
	\end{cases}
\end{equation}
where $\Gamma(x)$ is the gamma function. For integer $n$ we note that $Z_{2n + 1}\left(x\right)$ is an even function and $Z_{2n}\left(x\right)$ is odd. The first few of these associated Fried and Conte integrals are
\begin{align}
	Z_1(x) &= 1 + x Z_0(x),\\
	Z_2(x) &= x + x^2 Z_0(x),\\
	Z_3(x) &= \frac{1}{2} + x^2 + x^3 Z_0(x),\\
	Z_4(x) &= \frac{x}{2} + x^3 + x^4 Z_0(x).
\end{align} 

We also define a further generalization of the Fried and Conte integral as described in \citet{gurcan2014}:
\begin{equation}
	G_m(x_1, x_2) = \frac{1}{\sqrt{\pi}} \int^{\infty}_{-\infty} dv \frac{v^m e^{-v^2}}{\left(v-x_1\right)\left(v-x_2\right)}
\end{equation}
Through partial fraction decomposition, we can rewrite this as
\begin{equation}
	G_m(x_1, x_2) = \frac{1}{\sqrt{\pi}} \int^{\infty}_{-\infty} dv \left(\frac{1}{v-x_1} + \frac{x_2}{\left(v-x_1\right)\left(v-x_2\right)}\right)v^{m-1}e^{-v^2} = Z_{m-1}(x_1) + x_2 G_{m-1}(x_1, x_2). 
\end{equation}
Because $G_m(x_1, x_2) = G_m(x_2, x_1)$, we obtain
\begin{equation}
	G_m(x_1, x_2) = Z_{m-1}(x_1) + x_2 G_{m-1}(x_1, x_2) = G_m(x_2, x_1) =  Z_{m-1}(x_2) + x_1 G_{m-1}(x_1, x_2),
\end{equation}
which allows us to write 
\begin{equation}
	G_m(x_1, x_2) = \frac{Z_m(x_1) - Z_m(x_2)}{x_1 - x_2}. 
\end{equation}
Note that $G_m(x_1, x_2) = G_m(-x_1, -x_2)$. 

\section{Derivation of the Magnetic Drift Velocity} \label{Magnetic}
The goal of this section is to calculate the magnetic drift velocity $\vec{v}_{D, B}$ in the $s-\alpha$ equilibrium by including a finite Shafranov shift. We define the right-handed coordinate system $\left(r, \theta, \varphi\right)$ using Cartesian coordinates and include the Shafranov shift explicitly:
\begin{align}
x &= \left(R_0 + r \cos(\theta) + \Delta(r)\right) \cos(\varphi),\\
y &= \left(R_0 + r \cos(\theta) + \Delta(r)\right) \sin(\varphi),\\
z & = r \sin(\theta).
\end{align}
Here, $\Delta$ is the outward radial shift of the circular flux surface's center. The coordinate system $\left(r, \theta, \varphi\right)$ is right-handed but not orthogonal, so we must specify the metric coefficients. They are
\begin{align}
	g_{r r} & = 1 + \left(\Delta'\right)^2 + 2 \Delta' \cos\left(\theta\right),\\
	g_{r\theta} &= g_{\theta r} = - \Delta' r \sin\left(\theta \right),\\
	g_{\theta \theta} &= r^2,\\
	g_{\varphi \varphi} &= \left(R_0 + r \cos\left(\theta\right) + \Delta \right)^2,\\
	g_{r \varphi} &= g_{\varphi r} = g_{\theta \varphi} = g_{\varphi \theta} = 0,
\end{align}
where $\Delta' = \partial_{r} \Delta$. This leads to the Jacobian
\begin{equation}
	J = \sqrt{g} = \frac{1}{\nabla r \cdot \left(\nabla \theta \times \nabla \varphi\right)} =  r \left(R_0 + r \cos\left(\theta\right) + \Delta \right) \left(1 + \Delta' \cos\left(\theta\right)\right). 
\end{equation}
We next define a magnetic field for the $s-\alpha$ equilibrium. As an approximation, we ignore the poloidal magnetic field and only consider the toroidal magnetic. Thus, the magnetic field is
\begin{equation}
	\vec{B} \approx B_0 R_0 \nabla \varphi.
\end{equation}
This guarantees that the magnetic field strength is
\begin{equation}
	 B = \frac{B_0 R_0}{R_0  + r \cos\left(\theta\right) + \Delta} = \frac{B_0 R_0}{R},
\end{equation}
where $R= R\left(r, \theta\right)$. It is well known that one can obtain an approximate expression for $\Delta'$ from the Grad-Shafranov equation to lowest-order in $\epsilon$ \citep{connor1983, candy2009, linder2016}. The expression is
\begin{equation}
	\Delta' \approx - \alpha = q^2 \beta \frac{R_0}{P} \frac{dP}{dr}. 
\end{equation}

The next step is to calculate the magnetic drift velocity,
\begin{equation}
	\vec{v}_{D,B} = \frac{m}{e B} \left(v_\parallel^2 + \frac{v_\perp^2}{2}\right) \frac{\vec{B} \times \nabla B}{B^2} + \frac{m v_{\parallel}^2 }{e B} \frac{\beta}{2 p} \frac{\vec{B} \times \nabla p}{B}.
\end{equation}
The first term is the sum of the grad-$B$ drift as well as the dominant component of the curvature drift. The second term is the portion of the curvature drift that arises from considering the lowest-order MHD equilibrium. Since QuaLiKiz is applied in the regime where $\alpha$ is small, we ignore the second term entirely; this is equivalent to assuming that the magnetic field is approximately curl-free. Taking note that we are not using an orthogonal coordinate system, we find that the relevant cross product is
\begin{equation}
	\left(\vec{B} \times \nabla B\right) = R_0 B_0 \left(\frac{\partial B}{\partial r} \nabla \varphi \times \nabla r + \frac{\partial B}{\partial \theta} \nabla \varphi \times \nabla \theta\right). 
\end{equation}
We can evaluate each component of the expression to obtain
\begin{align}	
	\left(\vec{B} \times \nabla B\right) \cdot \nabla r & = R_0 B_0 \frac{\partial B}{\partial \theta} \nabla r \cdot \left(\nabla \varphi \times \nabla \theta\right) = -\frac{R_0 B_0}{J} \frac{\partial B}{\partial \theta} = -\frac{B^2 \sin\left(\theta\right)}{R \left(1 + \Delta' \cos\left(\theta\right)\right)} ,\\
	\left(\vec{B} \times \nabla B\right) \cdot \nabla \theta & = R_0 B_0 \frac{\partial B}{\partial r} \nabla \theta \cdot \left(\nabla \varphi \times \nabla r\right) = \frac{R_0 B_0}{J} \frac{\partial B}{\partial r} = -\frac{B^2\left(\cos\left(\theta\right) + \Delta'\right)}{R r \left(1 + \Delta' \cos\left(\theta\right)\right)} ,\\
	\left(\vec{B} \times \nabla B\right) \cdot \nabla \varphi &= 0.	
\end{align}
We then use the approximation
\begin{align}
	\frac{1}{R_0\left(1 + \Delta' \cos\left(\theta\right)\right)} \approx \frac{1}{R_0} \left(1 - \Delta' \cos\left(\theta\right)\right)
\end{align}
and substitute in $\Delta' = -\alpha$ to obtain to lowest-order
\begin{align}
	\vec{v}_D \cdot \nabla r & \approx - v_{D,B} \sin\left(\theta\right),\\
	\vec{v}_D \cdot \nabla \theta  & \approx - \frac{v_{D,B}}{r} \left(\cos\left(\theta\right) - \alpha \sin^2\left(\theta\right)\right),\\
	\vec{v}_D \cdot \nabla \varphi & \approx 0,
\end{align}
where we define the characteristic magnetic drift speed to be
\begin{equation}
	v_{D,B} = \frac{m}{eBR_0} \left(v_\parallel^2 + \frac{v_\perp^2}{2}\right).
\end{equation}

\section{Collisions} \label{Collisions}
The main sections of this work only consider the collisionless Vlasov equation. In actuality, QuaLiKiz implements a Krook-type collision operator for trapped electrons. To account for its inclusion, we modify the Vlasov equation to 
\begin{equation}
\frac{\partial \delta f_s}{\partial t} + \bm{\Omega} \cdot \frac{\partial \delta f_s}{\partial \bm{\alpha}} - e_s \frac{\partial \phi}{\partial \bm{\alpha}} \cdot \frac{\partial f_{0s}}{\partial \vec{J}} = -\nu \left(\delta f_s + \frac{e \phi}{T_s} f_{0s}\right),
\end{equation}
where $\nu$ is the collision frequency. Note that the $e_s \phi f_{0s} / T_s$ term accounts for the adiabatic response from the electrostatic perturbation. We only include this term for electron-ion collisions as ion-ion collisions and electron-electron collisions would produce only a small correction. Thus, we drop the ``s'' in favor of ``e'' and take $e_s \to -e$. Substituting in our Fourier expressions for $\delta f$ and $\phi$, we find that
\begin{equation}
f_{\vec{n}} = \frac{f_{0e}}{T_e} \frac{-e \phi_{\vec{n}}\left(\vec{n} \cdot \bm{\omega}_\ast - \vec{n} \cdot \bm{\Omega} - \nu\right)}{\vec{n} \cdot \bm{\Omega} - \overbar{\omega} - i \nu} = \frac{e \phi_{\vec{n}}}{T_e} f_{0e} \left(1 - \frac{\overbar{\omega} - \vec{n}\cdot \bm{\omega}_\ast}{\overbar{\omega} + i \nu - \vec{n} \cdot \bm{\Omega}}\right).
\end{equation}
Therefore, we can simply substitute $\omega \to \omega + i \nu$ in the denominator of the resonant term  to capture the effect of this collision operator. The drawback is that we lose the ability to simplify the functional. In QuaLiKiz, we take the collisional frequency to be
\begin{equation}
\nu_e(\xi, \lambda, \epsilon) = \nu_{ei} \left(\xi\right)^{-3/2} Z_{\text{eff}} \frac{\epsilon}{(1 - 2 \epsilon - \lambda)^2} \frac{0.111\delta + 1.31}{11.79 \delta + 1},
\end{equation}
$\nu_{ei}$ is the electron-ion Coulomb collision frequency, $Z_{\text{eff}}$ is the effective charge of the ion species interacting with the electrons, and the parameter $\delta$ is defined as
\begin{equation}
\delta = 12.0 \left(\frac{\left|\overbar{\omega}\right| \epsilon }{\nu_{ei} Z_{eff} }\right)^{3/2}.
\end{equation}
The explicit definition of $\nu_{ei}$ is
\begin{equation}
	\nu_{ei} = \frac{e^4 \lambda_e}{4 \pi \epsilon_0^2 \left(2 T_e\right)^{3/2} m_e^{1/2}},
\end{equation}
where $\lambda_e$ is the Coulomb logarithm relevant for electron collisions. Details for this collision operator can be found in \citet{romanelli2007}. The numerical values as well as the derivation of $\delta$ were originally calculated in \citet{kotschenreuther1995} and then modified for QuaLiKiz's purposes. Because $\nu$ is a function non-trivial of $\xi$, we cannot simplify the functional using this collision operator using Fried and Conte integrals, and the integration over the energy must be done numerically. The inability to simplify the $\xi$ integration results in $2$-dimensional integral. That aside, all other aspects of the trapped functional derivation remain intact.

We note this specific form of the collision operator is modified in comparison to the one found in \citet{romanelli2007}. It was found that the previous form of the collision operator led to incorrect predictions for density profiles when used in QuaLiKiz (coupled to integrated modeling suites) in highly collisional regimes. In response, numerical parameters in the Krook operator were tuned to linear simulations in GENE. In doing so, we keep unchanged the generic dependence and numerical parameters derived from fundamental principles unchanged \citep{stephens_inprep}.

\nocite{*}
\bibliographystyle{jpp}

\bibliography{QuaLiKizTutorial}
	
\end{document}